\documentclass[aps,twocolumn,showpacs,floatfix,amssymb]{revtex4}
\usepackage{graphicx}
\begin{document}

\title{Dynamic  spin susceptibility of
 superconducting cuprates:\\
 A microscopic theory of the magnetic resonance mode}

\author{ A.A. Vladimirov$^{a}$, D. Ihle$^{b}$, and N. M. Plakida$^{a,c}$ }
\affiliation{ $^a$Joint Institute for Nuclear Research, 141980
Dubna,
Russia\\
$^{b}$ Institut f\"{u}r Theoretische Physik, Universit\"{a}t
Leipzig, D-04109, Leipzig, Germany \\
 $^{c}$Max-Planck-Institut f\"{u}r Physik komplexer Systeme,
D-01187, Dresden,  Germany}

\date{\today}

\begin{abstract}
A microscopic theory of the dynamic  spin susceptibility (DSS) in
the superconducting state within the $t$--$J$ model is presented.
It is based on an exact representation for  the DSS obtained by
applying  the  Mori-type projection technique for the relaxation
function  in terms of  Hubbard operators. The static spin
susceptibility is evaluated  by  a sum-rule-conserving
generalized mean-field approximation, while the self-energy is
calculated in the mode-coupling approximation.  The spectrum of
spin excitations  is studied in a homogeneous phase of the
underdoped and optimally doped regions. The DSS reveals a
resonance mode (RM) at the antiferromagnetic wave vector ${\bf Q}
= \pi(1,1)$ at low temperatures  due to a strong suppression of
the damping of spin excitations. This is explained by an
involvement of spin excitations in the decay process besides the
particle-hole continuum usually considered in random-phase-type
approximations. The spin gap in the spin-excitation spectrum at
${\bf Q}$ plays a dominant role in limiting  the decay in
comparison with the superconducting gap which results in the
observation of the RM even above $T_c$ in the underdoped region.
A good agreement with inelastic neutron-scattering experiments on
the RM in YBCO compounds is found.
\end{abstract}


\maketitle

\section{Introduction}

In the superconducting state the spin-excitation spectrum of
high-$T_{\rm c}$ cuprates is dominated by a sharp magnetic peak
at the planar antiferromagnetic (AF) wave vector ${\bf Q} =
\pi(1,1)$ which is called {\it the resonance mode} (RM). It was
discovered  in the inelastic neutron  scattering (INS) experiments
which revealed a suppression of the spectral weight of low-energy
spin excitations at low  temperatures and its transfer  to higher
energies resulting in the RM. There is a vast literature devoted
to experimental and theoretical investigations of spin-excitation
spectra and  the RM in cuprates (a list of  references can be
found in the reviews
\cite{Bourges98,Sidis04,Sidis07,Eschrig06,Plakida10} and in a
number of publications, as  e.g.,
Refs.~\cite{Si93,Fong00,Dai01,Yamase06}). Here we consider only
the main results of these studies relevant for our purposes.
\par
The RM was discovered at first in the optimally doped
YBa$_{2}$Cu$_{3}$O$_{y}$~(YBCO$_y$) crystal at the energy $E_{\rm
r} \approx 41$~meV~\cite{Rossat91}  but later on, the RM was
found in the underdoped YBCO crystals,
Bi$_{2}$Sr$_{2}$CaCu$_{2}$O$_{8+\delta}$ (Bi-2212) compounds, and
other cuprates as well.~\cite{Sidis04} In particular, the RM was
observed in the single-layer cuprate superconductors
Tl$_2$Ba$_2$CuO$_{6+x}$ (Tl-2201) ~\cite{He02} and
HgBa$_2$CuO$_{4+\delta}$ (Hg-1201)~\cite{Yu10}, and in the
electron-doped Pr$_{0.88}$LaCe$_{0.12}$CuO$_{4-\delta}$
superconductor~\cite{Wilson06}. This demonstrates that the RM is
a generic feature of the cuprate superconductors and can be
related to spin excitations in a single CuO$_2$ layer. Since the
energy of the RM was found to scale with the superconducting
temperature, $E_{\rm r} \approx 5.3 k_{\rm B}T_{\rm c}$ in YBCO
and Bi-2212 compounds and $E_{\rm r} \approx 6\, (6.8) \,k_{\rm
B}T_{\rm c}$ in Tl-2201 (Hg-1201) systems, it has been argued
that it might constitute the bosonic excitation mediating
superconducting pairing in cuprates which has motivated an
extensive study of the RM phenomenon (see, e.g.,
Refs.~\cite{Eschrig06, Plakida10}).
\par
The spin-excitation dispersion close to the RM  exhibits  a
peculiar ``hour-glass''-like shape  with  upward and downward
dispersions. Whereas the RM energy $E_{\rm r}$  changes with
doping,  no essential temperature dependence of $E_{\rm r}$ and
of the upward branch of the dispersion has been found. In the
optimal doping region the RM and both dispersion branches are
smeared out above $ T_{\rm c}$. In the strongly underdoped YBCO
crystal only the downward branch is suppressed  above $ T_{\rm
c}$, whereas the upward dispersion and the RM are observed in the
normal pseudogap state. In particular, a well-defined resonance
peak at $E_{\rm r} \approx 33$~meV was found in the YBCO$_{6.5}$
crystal   in the oxygen ordered ortho-II phase with  $T_{\rm c} =
59$~K at hole doping $p = 0.09$ (Refs.~\cite{Stock04,Stock05}).
At low temperature, $T \sim 8$~K, the RM revealed a much higher
intensity than in optimally doped crystals, and it was also seen
with less intensity even at $T \simeq 1.4\,T_{\rm c}$.
\par
The extensive study of the twin-free YBCO$_{6.6}$ crystal    with
$T_{\rm c} = 61$~K at hole doping $p =
0.12$~(Refs.~\cite{Hinkov07,Hinkov10}) revealed a different origin
of the upward and downward parts of the spin  excitations close to
the RM energy $\omega_{\rm r} = 38.5$~meV. Specifically, the
high-energy excitations do not show  noticeable changes at the
superconducting transition and have the symmetry of the square
CuO$_2$ lattice common to all cuprates. Contrary to this, the
low-energy part of the spectrum is qualitatively different in the
superconducting and pseudogap states.  In Ref.~\cite{Hinkov10}
the INS data was fitted by a spectral function of spin excitations
in absolute units which was used to calculate the effective
spin-fluctuation-mediated pairing interaction and to estimate the
superconducting temperature $T_{\rm c}\sim 170$~K.~\cite{Dahm09}
\par
An hour-glass structure of the spin-excitation spectrum similar
to YBCO was found  in La$_{2-x}$Sr$_{x}$CuO$_4$ (LSCO-$x$)
crystals (see Refs.~\cite{Christensen04,Vignolle07,Lipscombe09}).
However, no sharp RM was found at $T < T_{\rm c}$ which may be
caused by strong disorder effects produced by Sr doping close to
the CuO$_2$ plane or by dynamically fluctuating stripes. A
double-peaked structure was observed in the local susceptibility
$\chi''(\omega)$  of the optimally doped LSCO-0.16
crystal~\cite{Vignolle07} with $T_{\rm c} = 38.5$~K. The
lower-energy incommensurate (IC) part of the spectrum was found
below $\omega \sim 20$~meV which may be explained by electron-hole
collective excitations (or fluctuating stripes), whereas the
high-energy upward dispersion,  peaked at $\omega = 40 - 50$ ~meV,
can be viewed as overdamped spin-wave-like excitations caused by
the residual AF exchange interaction  $J \sim 80$~meV.  A
similar, though less pronounced, double-peaked structure in
$\chi''(\omega)$ was found in the underdoped LSCO-$0.085$ crystal
with $T_{\rm c} = 22$~K  in the pseudogap
phase.~\cite{Lipscombe09}
\par
Generally, the INS experiments suggest the universality of  the
hour-glass structure of spin excitations in cuprates, though with
material dependent details.  The high-energy excitations above
the RM  can be considered as overdamped AF spin waves with the
exchange interaction decreasing with doping. Since the
characteristic exchange energy, especially at low doping, is quite
large, only a weak  temperature dependence of the spectrum is
observed resulting from the increase of the damping with
temperature that shifts the maximum of the dynamic  spin
susceptibility (DSS) to a lower energy. The low-energy spin
dynamics, which reveals a strong temperature dependence below
$T_{\rm c}$, may be related to fluctuating stripe phases with a
quasi-one-dimensional order of spins and charges or to a nematic
(liquid crystal) state  as discussed later.
\par
To explain the RM  in superconducting cuprates, various
theoretical models have been proposed. In a large number of
studies the Fermi-liquid model of itinerant electrons was assumed
and the DSS was calculated by the random phase approximation (RPA)
for the one-band Hubbard model or by taking the RPA
susceptibility with the AF  superexchange interaction $J({\bf
q})$ (see, e.g., Refs.~\cite{Norman00,Manske01,Eremin05}) In this
approach, the RM is considered as a particle-hole bound state,
usually referred to as a \mbox{spin-1} exciton.  The state is
formed below the continuum of particle-hole excitations which is
gapped at a threshold energy $\omega_c \leq 2\Delta({\bf q}^*)$
determined by the superconducting $d$-wave gap $2\Delta({\bf
q}^*)$ at a particular wave vector $\,{\bf q}^*\,$ on the Fermi
surface (FS).  In the Fermi-liquid  approach, the $d$-wave
symmetry of the gap and the shape of the  FS which should cross
the AF Brillouin zone (BZ) are essential in explaining the ${\bf
q}$- and $\omega$-dependence of the DSS and the RM.
\par
More complicated electronic  models were also considered, e.g.,
in Refs.~\cite{Si92,Si93,Zha93,Liu95,Kao00}, where an extensive
study of magnetic interactions  was performed for the extended
three-band Hubbard model with large Cu on-site correlation $U$.
Using the slave-boson representation and $1/N$ expansion ($N$ is
the spin degeneracy) the spectrum of spin excitations was
calculated both for the normal state and for the superconducting
$d$-wave state. It was stressed that a difference of the magnetic
neutron scattering in the LSCO and YBCO compounds could be
explained by  fine details of the band
structure.~\cite{Zha93,Kao00}  To explain the peculiarity of the
DSS and the RM in the two-layer YBCO compounds, a bilayer exchange
interaction was invoked (see, e.g., Refs.~\cite{Liu95,Eremin07}).
Usually the downward IC dispersion is well reproduced by fitting
the electron interaction while the intensity of the upward
dispersion appears to be too weak. To describe the underdoped
regime close to the insulating (and AF) state, where a model of
itinerant electrons cannot be justified, a phenomenological
spin-fermion model was used (see, e.g., Ref.~\cite{Eschrig06}).
\par
The INS study of the slightly overdoped Bi-2212
crystal~\cite{Fauque07} and of
Y$_{1-x}$Ca$_{x}$B$_2$Cu$_3$O$_{6+x}$ (Y-CaBCO)
compounds~\cite{Pailhes06} seems to support the spin-exciton
scenario. In particular, the global momentum shape of the
measured magnetic excitations is quantitatively  described within
the spin-exciton model with parameters inferred from
angle-resolved photo-emission experiments on Bi-2212 or
electronic Raman scattering experiments on Y-CaBCO. However, the
temperature dependence of the RM energy was not studied which
should be observed in the spin-exciton model due to the
temperature dependence of the superconducting gap.
\par
The strong correlation limit in the underdoped region in cuprates
is often treated within the $t$--$J$ model suitable for
consideration of low-energy spin dynamics. To take into account
the projected character of electron operators, the slave-boson
technique was used (for a review see Ref.~\cite{Ogata08}). In
particular, in Ref.~\cite{Brinckmann02} the magnetic excitation
spectrum was studied within the $t$-$t'$-$J$  model using  the
mean-field slave-particle theory. The DSS was calculated in the
RPA for spinons in the superconducting and spin-gap states. The
RM was revealed as the  spin-1 exciton below the threshold energy,
as in the Fermi-liquid models. A qualitative agreement with INS
experiments on YBCO and Bi-2212 compounds was obtained.  In
Ref.~\cite{Yamase06}, a comprehensive analysis of the DSS within
the mean-field slave-boson theory for the bilayer $t$--$J$ model
was  performed. A special emphasis was put on the explanation of
an anisotropy of the spectrum in the square CuO$_2$ plane which
was related to a $d$-wave shaped FS deformation ($d$FSD). In the
case of a strong $d$FSD a spontaneous breaking of the
orientational symmetry of the FS can occur  leading to its
orthorhombic deformation (Pomeranchuk instability). The DSS was
calculated in the RPA with a renormalized in-plane AF exchange
interaction. Within the theory, both the collective RM and its
downward dispersion  as well the high-energy excitations with the
upward dispersion were obtained. It was also possible  to explain
the in-plane anisotropy of the DSS observed in INS experiments by
the dFSD effect.
\par
However, in the  slave-particle  theory one has to introduce a
local constraint at each lattice site to reduce the enlarged
Hilbert space   (four states per site) to the physical one of the
projected electronic states (three states per site in the
$t$--$J$ model). In the mean-field approximation  the local
constraint is relaxed and substituted by an averaged one which
violates the  constraint and leads to uncontrollable results.
\par
A  rigorous approach is based on the Hubbard operators (HOs)
acting in the correct physical space~\cite{Hubbard65}. In
Ref.~\cite{Izyumov91}, a special diagram technique for the HOs was
developed  and used to calculate the DSS. In particular, a
generalized RPA was elaborated  by summing up  different  types
of bubble diagrams. However, the RM excitation was not considered
there. Calculations of the DSS within the HO technique in the
conventional RPA in Ref.~\cite{Onufrieva02} revealed a sharp RM,
caused by the opening of the superconducting gap below $T_{\rm
c}$, and low-energy collective excitations similar to the
Fermi-liquid models.
\par
Let us make a general remark  concerning the \mbox{spin-1}
exciton scenario based on a summation of fermion-bubble-type
diagrams, as in the conventional RPA. In this scenario, a strong
temperature dependence of the RM energy is expected below $T_{\rm
c}$ due to the temperature dependence of the superconducting  gap
$2\Delta({\bf q}^*)$ and, hence, of the threshold energy
$\omega_c$ and $E_{\rm r}< \omega_c$. However, this contradicts
to experiments. In theoretical calculations, usually only the
low-temperature limit, $T \ll T_{\rm c}$, and the $T > T_{\rm c}$
region are analyzed. It would be important to study a region at
$T \leq T_{\rm c}$ to test  the  \mbox{spin-1} exciton scenario.
\par
To go beyond a simple RPA in the strong correlation limit, the
Mori projection technique~\cite{Mori65} in the equation of motion
method for the relaxation function has been used by several
groups (see, e.g., Refs.
\cite{Sega03,Prelovsek04,Sega06,Prelovsek06,Sherman03,Sherman06}).
In the $t$--$J$ model, this method  affords to consider   the
magnetic excitations of localized spins in the undoped case
within the AF Heisenberg model and a crossover to the itinerant
electron spin excitations in  the overdoped region. These studies
and, in particular, the DSS calculation and the RM analysis in the
superconducting state will be discussed later and compared with
our results. In Refs.~\cite{Vladimirov05} and
\cite{Vladimirov09},  we have formulated a rigorous theory of the
DSS in the normal state within the projection operator method for
the relaxation function  in terms of the HOs. The obtained
results, both for the static properties (like the staggered
magnetization at $T=0$, the uniform static susceptibility, and
the AF correlation length) and for the DSS (e.g., the
$\,(\omega/T)$-scaling  behavior of the local DSS) have shown a
good agreement with available cluster calculations and
neutron-scattering  data. In particular, it was shown that the
kinematical interaction resulting from  the HO commutation
relations plays an essential role and gives the major
contribution to the damping of spin excitations induced by the
hopping term $t$.
\par
In this respect, we  mention a calculation of the DSS using the
Mori projection technique for the $t$--$J$ model.\cite{Zeyher10}
In this study, the projected character of electron operators was
neglected by replacing them with the conventional Fermi
operators. As a result,  the kinematical interaction induced by a
large  hopping term $t \simeq 3 J$ was disregarded and only the
exchange interaction $J$ was considered. The resulting Mori
memory function was represented as a sum of two contributions,
one determined by electrons  in the conventional RPA  and the
second given by the spin-spin scattering $\propto J^2$.  Though
the upward and downward dispersions of the spin-excitation
spectrum and the RM below $T_{\rm c}$ were reproduced, ignoring
the hopping term contribution to the memory function  can give
only a qualitative description.
\par
The discovery of the static IC charge- and spin-density waves (CDW
and SDW), referred to as stripes, in the metallic phase of
La$_{2-x}$Ba$_{x}$CuO$_4$ (LBCO-$x$) and LSCO compounds  has
attracted much attention in the recent years and was used in the
explanation of the IC spin-excitation spectrum  in hole-doped
cuprates (for reviews see
Refs.~\cite{Tranquada07,Kivelson03,Vojta09}). In particular, in
the stripe phase of the LBCO-$0.125$ compound the INS study at $T
= 12$~K ($> T_{\rm c}$)  revealed high-energy spin excitations at
$ \omega > 40$~meV and a low-energy branch similar to spin waves
in the two-leg-ladder spin model.~\cite{Tranquada04} Various
stripe models  have been proposed to describe the experimental
data. In particular, spin-only models for the bond-centered
stripe with long-range magnetic order in Ref.~\cite{Vojta04} and
a two-dimensional (2D) model of coupled two-leg spin ladders in
Ref.~\cite{Uhrig04} were proposed which fitted quite well the
experimental data.  In Ref.~\cite{Seibold05}, both bond- and
site-centered stripes were considered within the time-dependent
Gutzwiller approximation for the  Hubbard model. Static stripe
spin and charge  order coexisting with the $d$-wave
superconductivity was studied within an extended Hubbard model in
the mean-field approximation  and RPA for the DSS in
Ref.~\cite{Andersen05}. Using  quantum Monte Carlo simulations a
detailed study of  magnetic excitations was performed for coupled
spin ladders.~\cite{Andersen07} Magnetic ordering both
perpendicular  and parallel to the stripe direction was found in
an array of antiferromagnetically coupled doped and undoped
two-leg ladders.~\cite{Konik08}
\par
However, the static stripes have not been detected in moderately
doped YBCO crystals, where only dynamically fluctuating stripes
can be suggested (see, e.g.,
Refs.~\cite{Dai01,Stock04,Stock05,Hinkov04,Hinkov07}). Moreover,
instead of a rigid stripe array proposed in Ref.~\cite{Mook00},
the 2D character of the IC spin fluctuations was observed in the
untwinned YBCO$_{6+x}$ crystals with $\, x = 0.6$ and
$0.85$.~\cite{Hinkov04} The IC peaks exhibiting the in-plane
anisotropy  were explained as a result of a possible
liquid-crystalline stripe phase or a nematic state. Therefore,
though the inhomogeneous stripe picture can be applied for LSCO
and LBCO systems, it  seems to be not a universal origin of the
hour-glass spin-excitation dispersion and the RM in cuprate
superconductors. Inhomogeneous phases of cuprates related to
fluctuating stripes  with a quasi-one-dimensional order of spins
and charges or to a liquid crystal state may be important only in
the explanation of the low-energy collective spin fluctuations.
\par
To clarify some of the open problems in describing the RM
phenomenon, such as the appearance of the RM above $T_{\rm c} $
and its weak temperature dependence, in the present paper  we
extend our microscopic  theory (Refs.~\cite{Vladimirov05}  and
~\cite{Vladimirov09}) to the superconducting state.  Although our
general formulation for the DSS is similar to the original Mori
memory function approach used in other publications, as in
Ref.~\cite{Sega03}, in the previous studies of the $t$--$J$ model
only  the bubble-type diagrams similar to the RPA were considered
which ignores the important role of spin excitations in the
decay  process. The energy gap at the AF wave vector ${\bf Q} $
of the order of the RM energy $E_{\rm r}$ in the spin-excitation
spectrum strongly reduces the damping at low temperatures, $T \ll
E_{\rm r} \simeq 5\, k_{\rm B}T_{\rm c}$, which results in the
emergence of a sharp peak in the spectral function. In the low
doping region, where the damping is extremely small, the RM is
found even above $T_{\rm c}$.  In the overdoped region, at hole
concentration $\delta \sim 0.2$ and high $T_{\rm c}$, the
spin-excitation damping becomes large  and the opening of the
superconducting gap enhances the intensity of the RM, so that it
becomes observable only below $T_{\rm c}$. So, as compared with
the spin-exciton scenario, we propose an alternative explanation
of the RM and the upper branch of the dispersion  which are
driven by the spin gap at ${\bf Q}$ instead of the
superconducting gap $2\Delta$. A good agreement of our results
for the temperature and doping dependence of the spin-excitation
spectrum and the RM with INS experiments provides a strong
support for the proposed theory. We have not found a lower branch
of the hour-glass spectrum which may be related to inhomogeneous
states in the CuO$_2$ plane neglected in our analysis.
\par
In the next section we present the basic formulas for the DSS and
the self-energy  which are a generalization of our theory in
Ref.~\cite{Vladimirov09} to the superconducting state. Numerical
results for the spin-excitation spectra are given in Sec.~III,
where the temperature and doping dependence of the damping and
the RM   are discussed.  The conclusion is given in Sec.~IV.
Details of the calculations within the mode coupling
approximation (MCA) are presented in the Appendix.

\section{Relaxation-function theory}

\subsection{Dynamic spin susceptibility}

It is convenient to consider the $t$--$J$ model in the Hubbard
operator  representation
\begin{eqnarray}
H &=& - \sum_{i \neq j,\sigma}t_{ij}X_{i}^{\sigma
0}X_{j}^{0\sigma}
 - \mu \sum_{i \sigma} X_{i}^{\sigma \sigma}
\nonumber \\
 &  + &\frac{1}{4} \sum_{i \neq j,\sigma} J_{ij}
\left(X_i^{\sigma\bar{\sigma}}X_j^{\bar{\sigma}\sigma}  -
   X_i^{\sigma\sigma}X_j^{\bar{\sigma}\bar{\sigma}}\right),
\label{b1}
\end{eqnarray}
where  $t_{ij}$ is the hopping integral and $J_{ij}$ is the
exchange interaction.   The Hubbard operators
$\,X_{i}^{\alpha\beta}=|i,\alpha\rangle\langle i,\beta| \,$
describe transitions between  three possible states at a site $i$
on a square lattice:  an empty state
$|i,\alpha\rangle=|i,0\rangle$ and  a singly occupied state
$|i,\alpha\rangle=|i,\sigma\rangle$ with spin $\sigma = \pm
(1/2), \;( \bar{\sigma} = - \sigma) $. The number and spin
operators in terms of the Hubbard operators read:
\begin{equation}
N_{i}=\sum_{\sigma}X_{i}^{\sigma\sigma}, \quad, S_{i}^{\sigma} =
X_{i}^{\sigma\bar{\sigma}},\; S_{i}^{z} =  \sum_{\sigma} \sigma
X_{i}^{\sigma\sigma} .
 \label{b1a}
\end{equation}
The  Hubbard operators obey the completeness relation $ X_{i}^{00}
+ \sum_{\sigma} X_{i}^{\sigma\sigma} = 1$  which preserves
rigorously, contrary to the slave-boson approach, the constraint
of no double-occupancy of any lattice site.  The Hubbard operators
have the  commutation relations $\, \left[X_{i}^{\alpha\beta},
X_{j}^{\gamma\delta}\right]_{\pm}=
\delta_{ij}\left(\delta_{\beta\gamma}X_{i}^{\alpha\delta}\pm
\delta_{\delta\alpha}X_{i}^{\gamma\beta}\right) \,$  which
results in the {\it kinematical interaction}. Here, the upper sign
pertains to Fermi-type operators like $X_{i}^{0\sigma}$ changing
the number of electrons, and the lower sign pertains to Bose-type
operators, such as the number operator or the spin operators,
Eq.~(\ref {b1a}). The chemical potential $\mu$ is  determined
from the equation for the average electron density   $ n =
\langle N_{i} \rangle = 1- \delta  $, where $\delta = \langle
X_{i}^{00} \rangle $ is the hole concentration.
\par
In Ref.~\cite{Vladimirov05},  applying the  Mori-type projection
technique~\cite{Mori65}, elaborated for the relaxation function,
we have derived an exact representation  for the DSS $\,
\chi({\bf q}, \omega) \,$ related to the retarded commutator Green
function (GF) (see Ref.~\cite{Zubarev60}),
\begin{equation}
\chi({\bf q}, \omega) = -\langle \!\langle {S}_{\bf q}^{+}|
{S}_{-\bf q}^{-} \rangle \!\rangle_{\omega} =  \frac{m({\bf q})} {
\omega_{\bf q}^2 +\omega \, \Sigma({\bf q},\omega) - \omega^2 } ,
 \label{b2}
\end{equation}
where $m({\bf q})=\langle [i\dot{S}^{+}_{\bf q}, S_{-\bf
q}^{-}]\rangle  = \langle [\, [{S}^{+}_{\bf q}, H], \, S_{-\bf
q}^{-}]\rangle$, and $\omega_{\bf q}$ is the spin-excitation
spectrum in a generalized mean-field approximation (GMFA). The
self-energy is given by the many-particle Kubo-Mori relaxation
function
\begin{equation}
\Sigma({\bf q},\omega)=[1/m({\bf q})]\, (\!( - \ddot{S}_{\bf
q}^{+}\,| - \ddot{S}_{-\bf q}^{-})\!)_{\omega}^{(\rm proper)}\, ,
 \label{b3}
\end{equation}
where    $- \ddot{S}_{\bf q}^{\pm} = [\, [{S}_{\bf q}^{\pm},H],
\,H] $ (for details see Ref.~\cite{Vladimirov05}). The Kubo-Mori
relaxation function and the scalar product are defined as (see,
e.g.,~Ref.~\cite{Tserkovnikov81})
\begin{equation}
(( A | B ))_{\omega}= - i \int_{0}^{\infty} dt  e^{i\omega t}
  (A(t), B),
   \label{b2a1}
\end{equation}
and
\begin{equation}
 (A(t), B) = \int_{0}^{\beta}d\lambda
   \langle A(t-i\lambda) B \rangle,
  \quad \beta = 1/k_{\rm B} T ,
\label{b2a}
\end{equation}
respectively.  The ``proper'' part of the relaxation function
(\ref{b3}) does not contain parts connected by a single zero-order
relaxation function  which corresponds to the projected time
evolution in the original Mori projection
technique~\cite{Mori65}. The spin-excitation spectrum is given by
the spectral function defined by the imaginary part of the DSS
(\ref{b2}),
\begin{eqnarray}
\chi''({\bf q}, \omega)  =  \frac{- \omega \, \Sigma{''}({\bf
q},\omega)\; m({\bf q})} {[\omega^2 -  \omega_{\bf q}^2 - \omega
\, \Sigma{'}({\bf q},\omega)]^2  + [\omega \, \Sigma{''}({\bf
q},\omega)]^2} \, ,
 \label{b4}
\end{eqnarray}
where $\, \Sigma({\bf q},\omega +i0^+)= \Sigma{'}({\bf q},\omega)
+ i \Sigma{''}({\bf q},\omega)$,  and $\,\Sigma{'}({\bf q},\omega)
= - \Sigma{'}({\bf q}, -\omega)$ and $ \Sigma{''}({\bf q},\omega)
=  \Sigma{''}({\bf q}, - \omega) < 0$ are the real and imaginary
parts of the self-energy, respectively.

\subsection{Static susceptibility}

The general representation of the DSS~(\ref {b2}) determines the
static susceptibility $\, \chi_{\bf q} = \chi({\bf q}, 0) \,$ by
the equation
\begin{equation}
\chi_{\bf q} = ({S}_{\bf q}^{+},S_{-{\bf q}}^{-}) =  m({\bf q})
/\omega_{\bf q}^2 \, .
 \label{b5}
\end{equation}
To calculate  the spin-excitation spectrum $\omega_{\bf q}$ the
equality
\begin{equation}
m({\bf q}) = (-\ddot{S}_{\bf q}^{+},S_{-{\bf q}}^{-}) =
 \omega_{\bf q}^2 \, ({S}_{\bf
q}^{+},S_{-{\bf q}}^{-}),
 \label{b6}
\end{equation}
is used, where the correlation function $(-\ddot{S}_{\bf
q}^{+},S_{-{\bf q}}^{-})$ is evaluated in the GMFA  by a
decoupling procedure  in the site representation  as described in
Ref.~\cite{Vladimirov09}. This procedure is equivalent to the MCA
for the two-time correlation functions. This results in the
spin-excitation spectrum
\begin{eqnarray}
\omega_{\bf q}^2 &=& 8t^2\lambda_1(1-\gamma_{\bf
q})(1-n-F_{2,0}-2F_{1,1})
\nonumber \\
 &+ & 4J^2(1-\gamma_{\bf q})
\big[\, \lambda_2\frac{n}{2}-\alpha_1C_{1,0}(4\gamma_{\bf q}+1)
\nonumber\\
&+&\alpha_2(2C_{1,1}+C_{2,0})\, \big],
 \label{b7}
\end{eqnarray}
where $t$ and $J$ are the hopping integral and the exchange
interaction for the nearest neighbors, respectively, and
$\gamma_{\bf q}=(1/2)\,(\cos q_x + \cos q_y)$ (we take the
lattice spacing $a$ to be unity). The static electron and spin
correlation functions are defined as
\begin{eqnarray}
  F_{n,m} \equiv F_{\bf R} &= &\langle X_{\bf 0}^{\sigma 0}\,
X_{\bf R}^{0\sigma}\rangle = \frac{1}{N}\sum_{{\bf q} }\,F_{\bf q}
 {\rm e}^{i {\bf q R}},
\label{b8a}\\
   C_{n, m}  \equiv C_{\bf R} &=& \langle S^-_{\bf 0} \,
   S^+_{\bf R} \rangle = \frac{1}{N}\sum_{{\bf q} }\,C_{\bf q}
 {\rm e}^{i {\bf q R}} ,
 \label{b8b}
\end{eqnarray}
where ${\bf R} = n {\bf a}_x + m {\bf a}_y \,$. The  GMFA
spectrum (\ref {b7}) is calculated self-consistently by using the
GMFA approximation for the static correlation function
(\ref{b8b}),
\begin{equation}
C_{\bf q} = \frac{m({\bf q})}{ 2 \, \omega_{\bf q}}\,\coth
\frac{\beta \, \omega_{\bf q}}{2} \,  .
  \label{n3}
\end{equation}
The decoupling parameters $\alpha_1, \alpha_2$ and $ \lambda_1,
\lambda_2$ in Eq.~(\ref {b7}) take into account the vertex
renormalization for the spin-spin and electron-spin interaction,
respectively, as explained in Ref.~\cite{Vladimirov09}. In
particular, the parameters $\alpha_1, \alpha_2$ are evaluated from
the results for the Heisenberg model at $\delta =0$ and are kept
fixed for $\delta \neq 0$. The parameters $ \lambda_1, \lambda_2$
are calculated from the sum rule $C_{0,0} = \langle S^+_0
S^-_0\rangle  = (1/2) (1- \delta)$ with a fixed ratio
$\lambda_1/\lambda_2 = 0.378$. In the superconducting state, the
electron correlation function $\,F_{\bf R} \,$ is calculated by
the spectral function for electrons  in the superconducting state
(see Eq.~(\ref{n1a})). The variation of the parameters $
\lambda_1, \lambda_2$ below  the superconducting transition  is
negligibly small and practically has no influence on the spectrum
$\omega_{\bf q} $.
\par
The direct calculation of $m({\bf q})$ yields
\begin{equation}
m({\bf q})= -8 \, (1-\gamma_{\bf q})\, \left[ t \, F_{1,0}+ J \,
C_{1,0} \right].
 \label{b9}
\end{equation}
Thus, the  static susceptibility (\ref {b5}) is explicitly
determined by Eqs.~(\ref {b7}) and (\ref {b9}).

\subsection{Self-energy}

In what follows, we consider the $t$-$J$ model at a finite hole
doping $\delta >  0.05$ when,  as discussed in
Ref.~\cite{Vladimirov09}, the largest contribution to the
self-energy (\ref {b3}) is $\Sigma_t({\bf q},\omega)$ coming from
the spin-electron scattering. It is determined   by the hopping
term  $H_t$ in the $t$-$J$ model according to the equation for the
spin-density operators: $- \ddot{S}_{\bf q}^{\pm} = [\, [{S}_{\bf
q}^{\pm},H_t], \,H_t] $. As described in the Appendix,  in the
MCA this contribution reads
\begin{eqnarray}
 && \Sigma''_t({\bf q},\omega)=
-\frac{\pi(2t)^4 (e^{\beta \omega }-1)}{ m({\bf q})\,\omega}
\int\int\int_{-\infty}^{\infty} d\omega_1  d\omega_2 d\omega_3
 \nonumber\\
&&\frac{1}{N^2}\sum_{{\bf q_1, q_2}}N(\omega_2) [1-
n(\omega_1)]n(\omega_3) \delta(\omega + \omega_1 - \omega_2 -
\omega_3 )
\nonumber\\
&&B_{\bf q_2}(\omega_2)
 \big[({\Lambda}^2_{\bf q_1, q_2, q_3} +
{\Lambda}^2_{\bf q_3, q_2, q_1})\, A^N_{\bf q_1}(\omega_1) \,
A^N_{\bf q_3}(\omega_3)
\nonumber\\
&&- 2 {\Lambda}_{\bf q_1, q_2, q_3} {\Lambda}_{\bf q_3, q_2,
q_1}\, A^S_{{\bf q_1} \sigma}(\omega_1)\, A^S_{{\bf q_3}
\sigma}(\omega_3) \big] ,
 \label{b10}
\end{eqnarray}
where ${\bf q}_3 = {\bf q}-{\bf q}_1-{\bf q}_2$. The Fermi and
Bose functions are denoted by $n(\omega)=
(e^{\beta\omega}+1)^{-1}$ and
$N(\omega)=(e^{\beta\omega}-1)^{-1}$. The vertex function
$\Lambda_{\bf q_1, q_2, q_3}$ is defined by Eq.~(\ref {A5}). Here
we introduced the  spectral functions:
\begin{eqnarray}
 A^N_{\bf q}(\omega)& =& -(1/\pi){\rm Im} \langle\langle
X^{0\sigma}_{\bf q}|X^{\sigma0}_{\bf q}\rangle\rangle_{\omega},
\label{b11a}\\
 A^S_{{\bf q} \sigma}(\omega)& =& -(1/\pi){\rm Im} \langle\langle
X^{0\sigma}_{\bf q}|X^{0 \bar\sigma}_{-\bf q}
\rangle\rangle_{\omega},
\label{b11b}\\
B_{\bf q}(\omega) &= &(1/\pi)\,\chi''({\bf q}, \omega),
 \label{b11c}
\end{eqnarray}
where $A^{N, S}_{\bf q}(\omega)$ are determined by the  retarded
anticommutator GFs for electrons (see Ref.~\cite{Zubarev60}). In
comparison with the expression for the self-energy in the normal
state considered in Ref~\cite{Vladimirov09}, in Eq.~(\ref{b10})
there is the contribution  proportional to  the anomalous GF
 $\, \langle\langle X^{0\sigma}_{\bf q}|X^{0
\bar\sigma}_{-\bf q} \rangle\rangle_{\omega}\,$ which is nonzero
in the superconducting state only.
\par
It should be emphasized that  the self-energy (\ref{b10}) is
determined by the decay of a spin excitation with the energy
$\,\omega\,$ and wave  vector ${\bf q}$ into   three excitations:
a particle-hole pair  and a spin excitation. This process is
controlled by the energy and momentum conservation laws, $\omega =
(\omega_3 - \omega_1) + \omega_2\,$ and $\, {\bf q} = {\bf q}_1 +
{\bf q}_2 + {\bf q}_3  $, respectively.  In  previous studies of
the $t$-$J$ model the contribution of the additional spin
excitation has been neglected (see, e.g.,
Ref.~\cite{Onufrieva02}) or approximated by static or
mean-field-type expressions (see, e.g., Refs.~\cite{Sega03} and
~\cite{Sherman03}). That is, in these approximations the
spin-excitation contribution was ``decoupled'' from the
particle-hole pair. We can derive the particle-hole bubble
approximation from Eq.~(\ref {b10}), if  we ignore the
spin-energy contribution $\omega_2$ in comparison with the
electron-hole pair energy, or, equivalently, if in the MCA,
Eqs.~(\ref{A6})  and (\ref{A7}), the time-dependent spin
correlation function is approximated by its static value:
$\langle S_{\bf -q}^{-} S_{\bf q}^{+}(t)\rangle \simeq \langle
S_{\bf -q}^{-} S_{\bf q}^{+}\rangle = C_{\bf q}$. Moreover,
excluding the spin-excitation wave-vector  ${\bf q}_2$ from the
wave-vector conservation law, we have ${\bf q} = {\bf q}_1 + {\bf
q}_3 $. As a result of these approximations in Eq.~(\ref {b10}),
we obtain the self-energy in the form of the particle-hole bubble
approximation:
\begin{eqnarray}
&&\widetilde{\Sigma}_t''({\bf q},\omega)
 =  -\frac{\pi(2t)^4\, }
{ m({\bf q})\,\omega} \int_{-\infty}^{\infty} d\omega_1 [
n(\omega_1)- n(\omega_1 +\omega)]
 \nonumber\\
&&  \times\frac{1}{N}\,\sum_{{\bf q_1}}\,
 \big[ \widetilde{\Lambda}^N_{{\bf q_1, q -q_1}}\, A^N_{\bf
q_1}(\omega_1) \, A^N_{\bf q - q_1}(\omega_1 +\omega)
\nonumber \\
&&  - \widetilde{\Lambda}^S_{{\bf q_1, q -q_1}}\, A^S_{{\bf q_1}
\sigma}(\omega_1)\, A^S_{{\bf q- q_1} \sigma}(\omega_1 +\omega)
\big]  ,
 \label{b12}
\end{eqnarray}
where the  averaged over the spin-excitation wave vector ${\bf
q_2}$ vertexes are introduced,
\begin{eqnarray}
 \widetilde{\Lambda}^N_{{\bf q_1, q_3}}& = & \frac{1}{N}
  \sum_{{\bf q_2}} \, C_{{\bf q}_2}
 [\Lambda^2_{\bf q_1, q_2, q_3} + \Lambda^2_{\bf q_3, q_2,
   q_1}] \, ,
\label{b12n} \\
\widetilde{\Lambda}^S_{{\bf q_1, q_3}} &=& \frac{2}{N}
 \sum_{{\bf q_2}} \, C_{{\bf q}_2} {\Lambda}_{\bf q_1, q_2, q_3}
{\Lambda}_{\bf q_3, q_2, q_1} .
 \label{b12s}
\end{eqnarray}
In the approximation (\ref{b12})  only the opening of  a
superconducting gap in the particle-hole excitation can suppress
the damping of spin excitations due to the decay into
particle-hole pairs which may result in the RM. Below we discuss
in more detail why  a particle-hole bubble approximation for the
self-energy, Eq.~(\ref {b12}), leads to  a different behavior of
the spin-excitation damping  in comparison with the results
obtained for the full self-energy (\ref{b10}).

\section{Results and Discussion}

\subsection{Self-energy approximation}

In the  calculation of the self-energy  (\ref {b10}) we adopt the
mean-field approximation (MFA) for the electron spectral functions
(\ref{b11a}) and (\ref{b11b})  which in the superconducting state
can be written as
\begin{eqnarray}
A^N_{\bf q}(\omega) & = & Q \sum_{\omega_1 = \pm E_{\bf
q}}\frac{\omega_1 + \varepsilon_{\bf q}}{2 \omega_1}
 \delta(\omega - \omega_1) \, ,
\label{n1a} \\
A^S_{{\bf q} \sigma}(\omega) &  = & Q \sum_{\omega_1 =\pm E_{\bf
q}}\frac{\Delta_{{\bf q} \sigma}} {2 \omega_1}
 \delta(\omega - \omega_1)\, .
\label{n1b}
\end{eqnarray}
Here $Q = 1 - n/2$ is the Hubbard weighting factor and the
superconducting gap function  $\, \Delta_{{\bf q} \sigma} = ({\rm
sgn} \;\sigma) \, \Delta_{\bf q}$. In the electron spectrum
$\varepsilon_{\bf q}$ we take into account only the
nearest-neighbor hopping $\,t \,$ and consider the energy
dispersion  in the Hubbard-I approximation: $\varepsilon_{\bf q}
= - 4t\,Q\, \gamma_{\bf q}- \mu$. The spectrum of quasiparticles
in the superconducting state is given by the conventional formula
$E_{\bf q}=\sqrt{\varepsilon_{\bf q}^2 + \Delta_{\bf q}^2} $. For
the spin-excitation  spectral function (\ref{b11c}) we take the
form:
\begin{equation}
B_{\bf q}(\omega) ={m({\bf q})}\sum_{\omega_1 = \pm
\widetilde{\omega}_{\bf q}}\frac{1}{2 \omega_1}   \delta(\omega -
\omega_1),
 \label{n2}
\end{equation}
where the spectrum of spin excitations $\widetilde{\omega}_{\bf
q}$  is determined by the pole of the DSS,
$\,\widetilde{\omega}_{\bf q} = [\omega_{\bf q}^2 +
\widetilde{\omega}_{\bf q} \, \Sigma'({\bf
q},\widetilde{\omega}_{\bf q})]^{1/2} \,$. Here, the  real part of
the self-energy $\Sigma'({\bf q},\omega)$ is calculated
perturbationally  by taking the spectral function (\ref{n2}) with
the GMFA spectrum $\omega_1 = \pm {\omega}_{\bf q}$. Using these
spectral functions, after integration over the energies in
Eq.~(\ref {b10}) we write the imaginary part of the self-energy
in the following form convenient for the calculation in the
limit  $T \to 0$:
\begin{eqnarray}
&&\Sigma''_t({\bf q},\omega)= \frac{\pi(2t)^4}{\omega m({\bf q})}
\frac{Q^2}{N^2}\sum_{{\bf q_1, q_2}} \sum_{\omega_1=\pm E_{{\bf
q}_1}}  \sum_{ \omega_2=\pm \widetilde{\omega}_{{\bf q}_2}
}\sum_{\omega_3=\pm E_{{\bf q}_3}}
\nonumber \\
&& m({\bf q}_2) \frac{N(\omega_2)n(-\omega_1) n(\omega_3) +
N(-\omega_2)n(\omega_1) n(-\omega_3)}
 {8\omega_1 \omega_2 \omega_3}
 \nonumber\\
&&\big[(\Lambda^2_{\bf q_1, q_2, q_3} + \Lambda^2_{\bf q_3,
q_2,q_1})
 (\omega_1+\varepsilon_{{\bf q}_1})(\omega_3+\varepsilon_{{\bf q}_3})
 \label{n4} \\
&& -2{\Lambda}_{\bf q_1, q_2, q_3}{\Lambda}_{\bf q_3, q_2, q_1}
\Delta_{{\bf q}_1 } \Delta_{{\bf q}_3 } \big]\delta(\omega
+\omega_1 -\omega_2 - \omega_3  ).
 \nonumber
\end{eqnarray}
Similar calculations for the  self-energy (\ref {b12}) in the
particle-hole bubble approximation  yield
\begin{eqnarray}
&&\widetilde{\Sigma}''_t({\bf q},\omega)= - \frac{\pi(2t)^4}{
m({\bf q})\omega} \frac{Q^2}{N}\sum_{{\bf q_1}}\,
\sum_{\omega_1=\pm E_{{\bf q}_1}} \, \sum_{\omega_2=\pm E_{{\bf q-
q}_1}}
 \nonumber \\
&&
 \frac{ n(\omega_1)-    n(\omega_2) }{4\omega_1  \omega_2}
  \big[\widetilde{\Lambda}^N_{{\bf q_1, q - q_1}}
 (\omega_1+\varepsilon_{{\bf q}_1})(\omega_2 +
 \varepsilon_{{\bf q - q}_1})
 \nonumber \\
&& -\widetilde{\Lambda}^S_{{\bf q_1, q - q_1}} \Delta_{{\bf q}_1
} \Delta_{{\bf q - q}_1 } \big] \, \delta(\omega +\omega_1
-\omega_2).\label{b12a}
\end{eqnarray}
\par
We consider both the $d$-wave and $s$-wave symmetry of the
superconducting  gap which we write as $\Delta^{(d)}_{\bf q} =
(\Delta/2)(\cos q_x - \cos q_y)$ and $\Delta^{(s)} = \Delta \, $
with the temperature dependent amplitude $\Delta(T)$.  In
numerical calculations we assume that  $\Delta(T)$ follows the
conventional Bardeen-Cooper-Schrieffer (BCS) theory. In
particular, $\Delta(T)/k_{\rm B} T_{\rm c} = 1.76, \, 1.72, \,
1.6, \, 1.24$ for  $T/T_{\rm c} = 0,\, 0.4,\, 0.6,\, 0.8$,
respectively. By taking, instead of the BCS ratio
$2\Delta_0/k_{\rm B} T_{\rm c} = 3.52,\, \Delta_0 = \Delta(T=0) $,
the ratio $ 2\Delta_0/ k_{\rm B} T_{\rm c} = 4.3 $ for a pure
$d$-wave superconductor (see, e.g., Ref.~\cite{Won94}) we have
found  that the results do not change noticeably, i.e., less then
$5\%$ at $T =0$ and even less at finite temperatures. We mainly
consider two doping values, $\delta=0.2$ which is  larger than
optimal doping and $\delta=0.09$ for the underdoped case. For
$\delta=0.2$ we fix the superconducting transition temperature as
$k_{\rm B} T_{\rm c} = 0.025 t$, while for $\delta = 0.09$ we
take $k_{\rm B} T_{\rm c} = 0.016 t$. For the hopping parameter
$t = 0.313$~eV these values are close to $T_{\rm c} = 91$~K in
the nearly optimally  doped YBCO$_{6.92}$ single crystal in
Ref.~\cite{Bourges98} and $T_{\rm c} = 59$~K in the underdoped
($\delta=0.09$) YBCO$_{6.5}$ crystal studied in
Ref.~\cite{Stock04}.  We take the exchange interaction $\,
J=0.3\,t $ and  measure all energies in units of $\, t \,$.

\subsection{Spin-excitation  damping}

\begin{figure}
\resizebox{0.35\textwidth}{!}{%
  \includegraphics{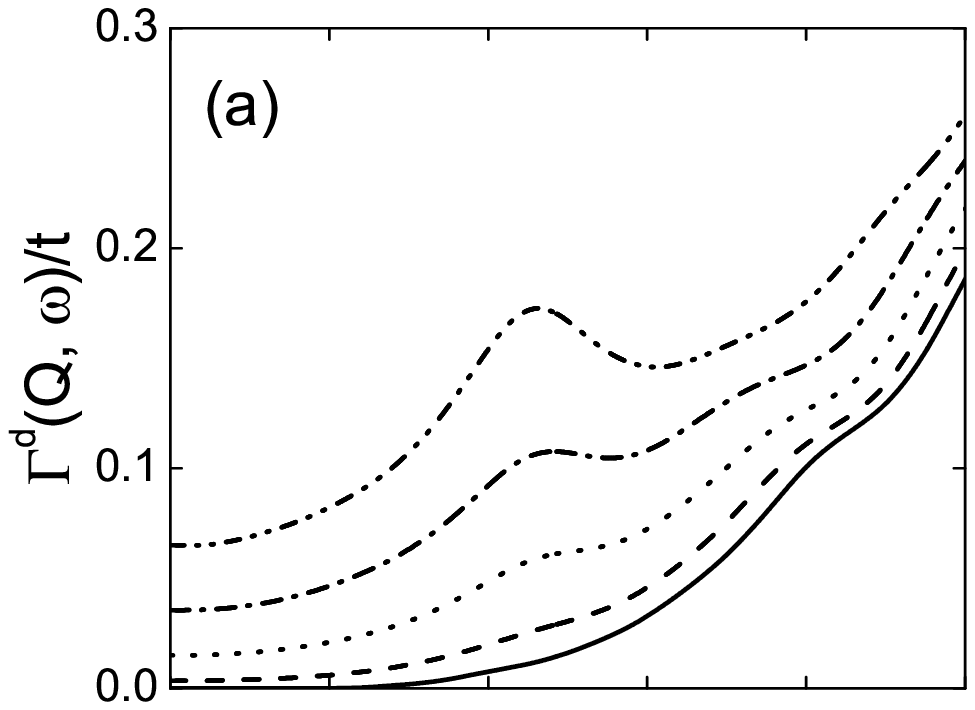}}\\
\resizebox{0.35\textwidth}{!}{%
\includegraphics{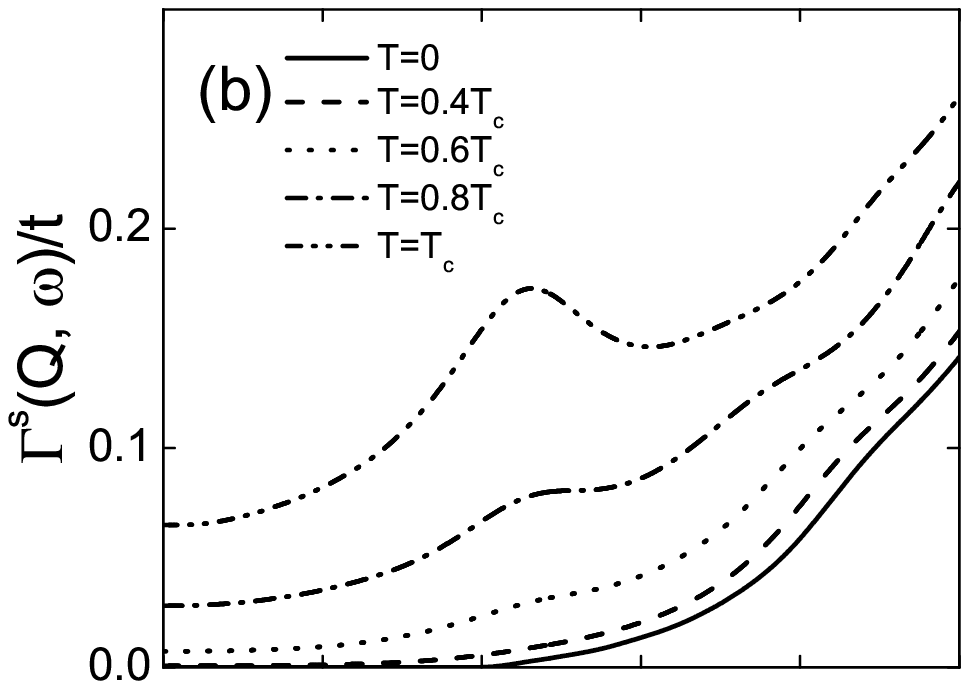}}\\
$\; \,$
\resizebox{0.365\textwidth}{!}{%
\includegraphics{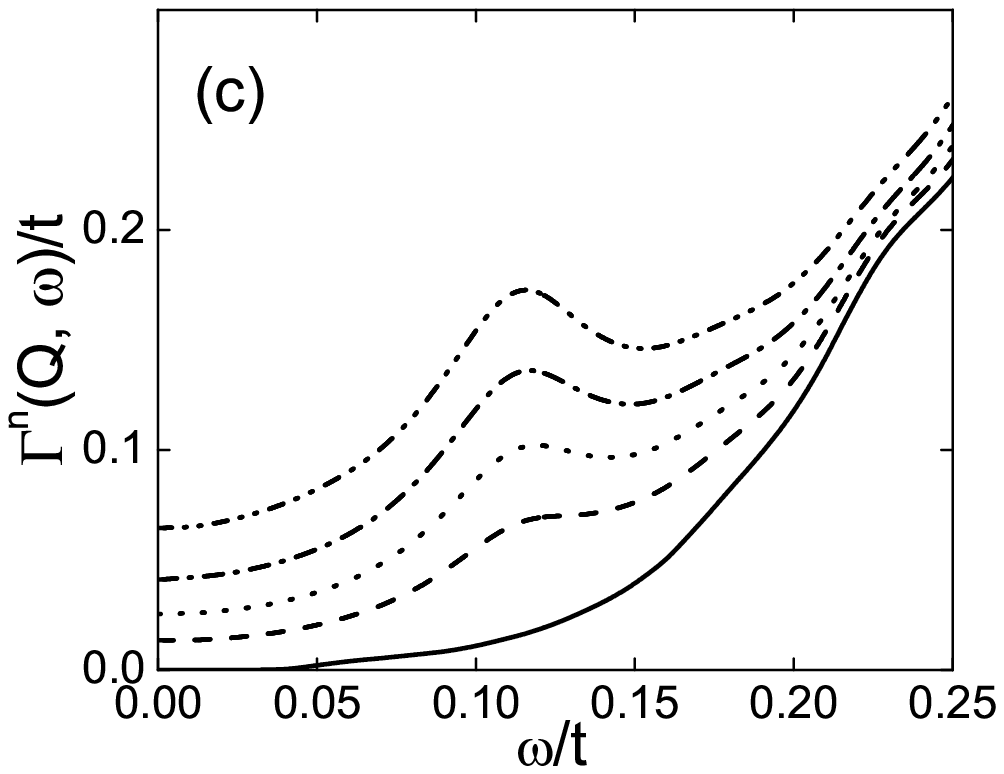}}
\caption{Spin-excitation damping $\Gamma({\bf Q},\omega)$ for
$\delta=0.2$ at $T \leq T_{\rm c}$ for (a) the $d$-wave and  (b)
$s$-wave pairing, and (c) in the normal state.}
 \label{Fig_1}
\end{figure}

To elucidate the role of  spin excitations in the damping and
their relevance to the shape of the spectral function,
Eq.~(\ref{b4}), we consider the temperature dependence of the
spin-excitation damping at the AF wave  vector, $\Gamma({\bf
Q},\omega) = - (1/2)\,\Sigma''_t({\bf Q},\omega) $.
Figure~\ref{Fig_1} shows the damping  for $\delta=0.2$ at various
temperatures  in  the case of  the $d$-wave (a) and  $s$-wave (b)
pairing. The difference of the damping appears only at low
$\omega$ and $T$.  In particular, the damping for the $s$-wave
gap at $T = 0$ disappears at $\omega < 2\Delta_0 \simeq 4 k_{\rm
B} T_{\rm c} = 0.1 t$, while for the $d$-wave gap it vanishes at
$\omega \simeq \Delta_0 \simeq  0.05 t$. A weak damping was
obtained also for the normal state shown in Fig.~\ref{Fig_1}~(c)
when the contribution from the superconducting gap functions in
the self-energy $\Sigma''_t({\bf Q},\omega)$, Eq.~(\ref {n4}), is
omitted. The similar smooth variation of the damping with  energy
in all three cases below $T_{\rm c}$, contrary to a step-like
dependence obtained in  the particle-hole bubble approximation
(see below), demonstrates that the superconducting gap plays a
minor role in suppressing the damping, and  the gap
$\widetilde{\omega}_{\bf Q}$ in the spin-excitation spectrum is
responsible for such a peculiar behavior.
\par
\begin{figure}
\resizebox{0.4\textwidth}{!}{%
  \includegraphics{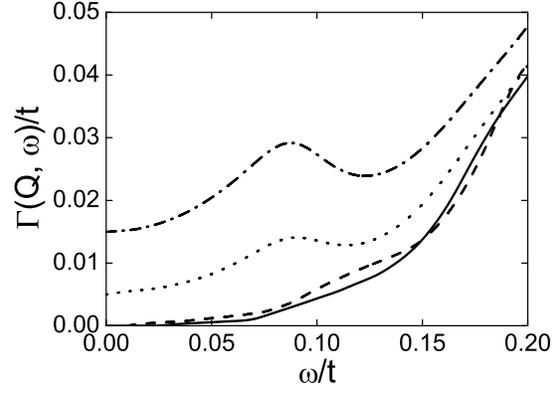}}
 \caption{Spin-excitation damping $\Gamma({\bf Q},\omega)$ for
$\delta=0.09$ for the $d$-wave pairing at $T = 0$ (solid line), $
T = T_{\rm c}$ (dotted line), and $ T = 1.4 T_{\rm c}$
(dash-dotted line).  The damping at $T = 0$ for the electron dispersion with
$t'/t = 0.37 $ and  $t'' / t = 0.1$  is shown by the dashed
line.}
 \label{Fig_2}
\end{figure}

For a lower doping, the damping becomes an order of magnitude
weaker, as shown in Fig.~\ref{Fig_2} at $\delta = 0.09$, even
above $T_{\rm c}\; (T =  1.4 T_{\rm c})$. For comparison with
results obtained in the fermion-bubble approximation, we
calculate  the damping also for a more general electron
dispersion, $\, \varepsilon_{\bf q} =  - 4 t \, \gamma_{\bf q} +
4 t' \,\gamma'_{\bf q} - 4 t''\, \gamma''_{\bf q}- \mu $, taking
into account hopping between next- and third-nearest neighbors,
$t'$ and $t''$,  respectively,  where $\, \gamma'_{\bf q} =
\,\cos q_x \cos q_y $  and $\gamma ''_{\bf q} = (1/2)(\cos 2 q_x
+\cos 2 q_y) $. In Fig.~\ref{Fig_2}  the damping at $T = 0$ for
the  parameters $ t'/t = 0.37$ and $t''/t = 0.1$,  proposed in
Ref.~\cite{Dahm09} for the  electron dispersion in the
antibonding band in YBCO$_{6.6}$, is shown by the dashed line.
The difference between the damping calculated for the electron
dispersion $\varepsilon_{\bf q} = - 4t\,Q\, \gamma_{\bf q}- \mu$
in Eqs.~(\ref{n1a}) and (\ref{n1b}) and the more general one with
nonzero parameters $t'$ and $ t''$ is negligible. This proves
that the shape of the FS and the electron dispersion are
unessential factors in our theory. To describe  the
electron-doped cuprates within the $t$--$J$ model,  we should
change the signs of the hopping parameters $t_{ij}$ in
Eq.~(\ref{b1}), in particular, $ - t \rightarrow  t > 0$. This
replacement does not change the obtained results and, therefore,
the theory can be also  applied to the electron-doped cuprates
with a proper fit of  model parameters.
\par
\begin{figure}
\resizebox{0.4\textwidth}{!}{%
  \includegraphics{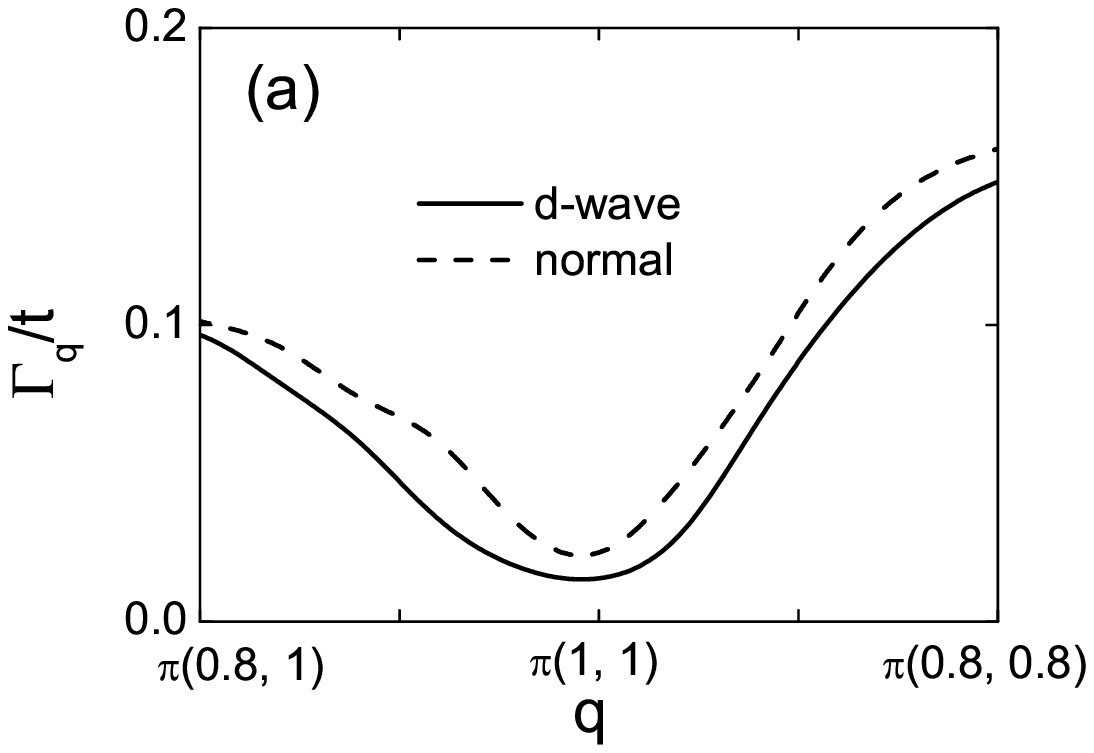}}\\
  \resizebox{0.41\textwidth}{!}{%
  \includegraphics{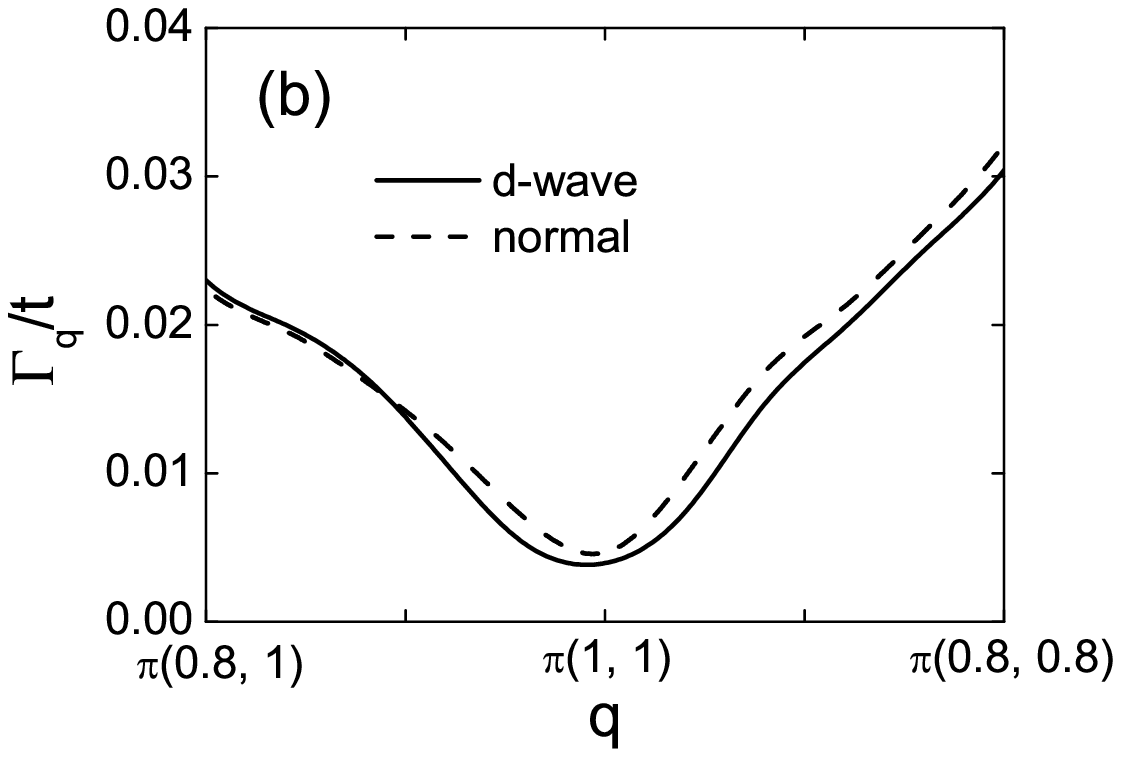}}
\caption{Spin-excitation damping $\Gamma_{\bf q}$ for (a)
$\delta=0.2$ and for (b)  $\delta=0.09$ at $T =0$ for  the
$d$-wave pairing (solid line) and   in the normal state (dashed
line). }
 \label{Fig_3}
\end{figure}

Sometimes,  the RM observed above $T_{\rm c}$ in the underdoped
cuprates is related to a pseudogap in the electronic spectrum
(see, e.g., Ref.~\cite{Eschrig06}). We  propose another
explanation: The RM above $T_{\rm c}$ is the result of the gapped
spin excitations in the self-energy (\ref {n4}) leading to  a very
weak damping in the underdoped region which is outlined in more
detail  below. This explanation  is supported by studies of the
spin-excitation damping $\Gamma_{\bf q} = -(1/2)
\,\Sigma''_t({\bf q},\omega = \widetilde{\omega}_{\bf q}) $  at
$T =0$ shown in Fig.~\ref{Fig_3}.  The small difference between
the damping in the $d$-wave superconducting state and the normal
state observed for the full self-energy, Eq.~(\ref{n4}), confirms
that the superconducting gap does not play an essential role in
suppressing  the damping $\Gamma_{\bf Q}$, in particular in the
underdoped region. At the same time, the sharp increase of
$\Gamma_{\bf q}$  away from the AF wave-vector ${\bf Q}$ explains
the resonance character of spin excitations at ${\bf Q}$.
\par
\begin{figure}
\resizebox{0.4\textwidth}{!}{%
 \includegraphics{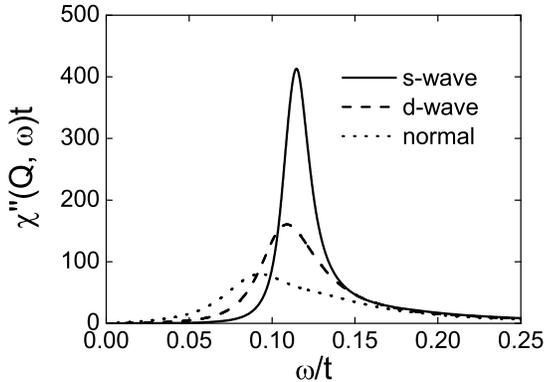}}
\caption{Spectral function $\chi''({\bf Q}, \omega)$ for the
$d$-wave and $s$-wave pairing in comparison with the normal state
at $T=0.4 T_{\rm c}$ for $\delta=0.2$.}
 \label{Fig_4}
\end{figure}

Though the damping in Fig.~\ref{Fig_1} looks similar, the spectral
functions shown in Fig.~\ref{Fig_4} for $\delta=0.2$  at  $T = 0.4
T_{\rm c}$ reveal a strong enhancement of the intensity of the RM
in the superconducting state.  In comparison to the normal state,
where the contribution from  the superconducting gap is omitted,
the peak intensity is about two (five) times larger for the $d
\,(s)$-wave  symmetry of the gap.
\par
\begin{figure}
\resizebox{0.4\textwidth}{!}{%
  \includegraphics{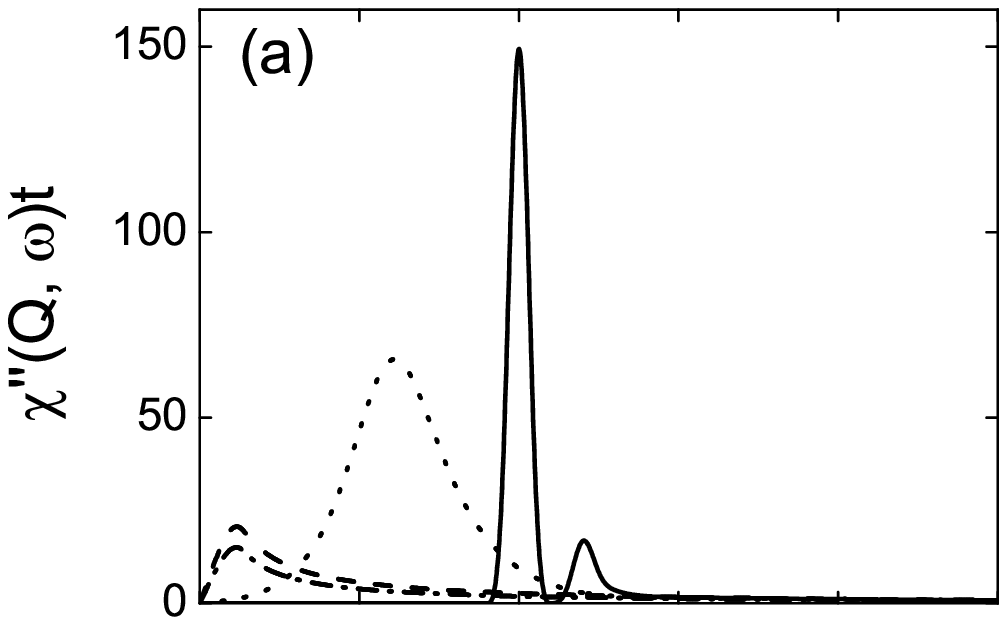}}\\
\mbox{$\;$} \resizebox{0.41\textwidth}{!}{%
\includegraphics{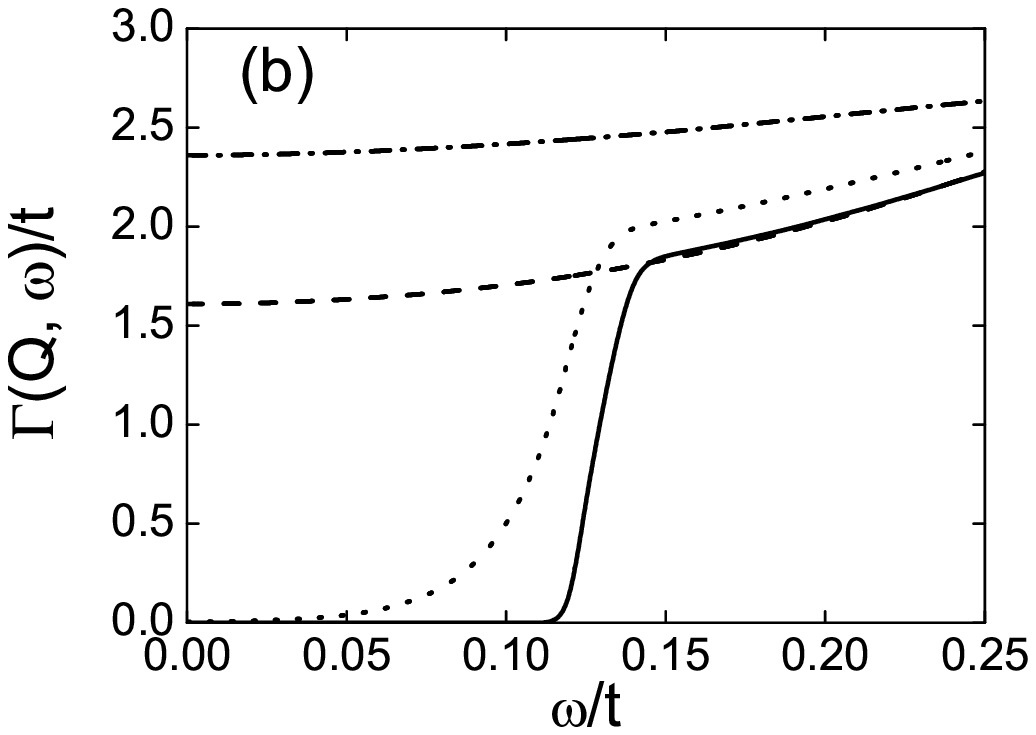}}
\caption{(a) Spectral function $\chi''({\bf Q}, \omega)$ and (b)
spin-excitation damping $\Gamma({\bf Q}, \omega)$ calculated   in
the particle-hole bubble approximation, Eq.~(\ref{b12}),  at
$\delta = 0.2$ for the $d$-wave pairing ($\Delta_0 = 0.1 t$ taken
from Ref.~\cite{Sega03} ) at $T = 0$ (solid line) and $T=0.4
T_{\rm c}$ (dotted line), and for the normal state at  $T = 0$
(dashed line) and $T= T_{\rm c}$ (dash-dotted line).}
 \label{Fig_5}
\end{figure}

Quite a different behavior of the damping and the spectral
function is obtained for the reduced self-energy,
Eq.~(\ref{b12a}), with a contribution only from a particle-hole
bubble. Figure~\ref{Fig_5} shows our  results for  the spectral
function $\,\chi''({\bf q}, \omega) \, $ and for the damping
$\Gamma({\bf Q}, \omega)$. To compare these functions with those
calculated in Ref.~\cite{Sega03}, we adopt  the electron
dispersion used in Ref.~\cite{Sega03}, $\,\varepsilon_{\bf
q}^{eff} = -4\,t_{eff} \gamma_{\bf q} - 4\, t'_{eff} \cos q_x
\cos q_y  \, $ with $ t_{eff} = 0.3t $ and $ t'_{eff} = - 0.1t\,$
and take the gap parameter $\, \Delta_0 = 0.1t$. The obtained
results are quite close to those  shown in Fig.~1 of
Ref.~\cite{Sega03} (where  $\chi_{zz}''({\bf q}, \omega) =
(1/2)\chi''({\bf q}, \omega)\,$ is plotted). At $T = 0$, we
observe a much narrower RM, but with a lower intensity  in
comparison with the RM calculated with the full self-energy,
Eq.~(\ref{n4}), as shown in Fig.~\ref{Fig_6}.  The energy $E_{\rm
r}$ of the RM shown in Fig.~\ref{Fig_5}~(a) noticeably decreases
with increasing temperature,  contrary to a negligible shift of
the RM shown in Fig.~\ref{Fig_6} for $T=0.4 T_{\rm c}$. This
comparison demonstrates that in the particle-hole bubble
approximation the superconducting gap plays a crucial role in the
occurrence of the RM with  $E_{\rm r}(T) < 2 \Delta(T)$, while in
the full self-energy  (\ref {n4}) the superconducting gap and
details of the electron dispersion  are less important. In
particular, for the underdoped case  $\delta = 0.09$ we have not
found  visible changes of the damping function shown in
Fig.~\ref{Fig_2} for the electron dispersion with $t' = 0$ and
$t' = - 0.1t$. For the reduced self-energy, Eq.~(\ref{b12a}), the
damping vanishes for both types of the dispersion in the
underdoped region, and  in order to obtain a finite  damping  at
$\delta = 0.1$ in Ref.~\cite{Sega03} (see Fig.~2),  the authors
have to use the electron density of states in the damping
function (see their Eq.~(20)) instead of the ${\bf q }$-dependent
electron spectral functions.
\par
This difference can be explained as follows. Whereas in the
particle-hole bubble approximation given by Eq.~(\ref{b12a}) the
spin excitation with the  energy $\omega$ at the wave vector ${\bf
Q}$  can decay only into a particle-hole pair with the energy
$\omega({\bf Q}) = E_{\bf Q + q} + E_{\bf q}$, in a more general
process described by Eq.~(\ref{n4}) an additional spin excitation
participates  in the scattering. In the limit $T \to 0$, the decay
process  is governed by another energy-conservation law,
$\omega({\bf Q}) = E_{\bf q_3} + E_{\bf q_1} +
\widetilde{\omega}_{\bf q_2}$, where the largest contribution from
the spin excitation comes from $\widetilde{\omega}_{\bf q_2}
\simeq \widetilde{\omega}_{\bf Q} $ due to the factor $\, m({\bf
q}_2)$, Eq.~(\ref{b9}), in Eq.~(\ref{n4}). This energy-momentum
conservation law strongly reduces the phase space for the decay
and suppresses the damping of the initial spin excitation with
the energy $\omega({\bf Q}) $. In fact, the occurrence of an
additional spin excitation   in the scattering process with the
finite energy $\widetilde{\omega}_{\bf Q}$ plays a role similar
to the superconducting gap in the excitation of the particle-hole
pair in  Eq.~(\ref{b12a}). Therefore, the damping at low
temperatures ($k_{\rm B}\,T \ll \widetilde{\omega}_{\bf Q} \sim
E_{\rm r} $) appears to be small even in the normal state as
demonstrated in Fig.~\ref{Fig_1}~(c). In the case of the
particle-hole relaxation, the condition for the occurrence of the
RM, $\,\omega({\bf Q}) = E_{\bf q + Q} + E_{\bf q} \leq 2
\Delta({\bf q}^*) $, imposes a strong restriction on the shape of
the FS which should cross the AF Brillouin zone to accommodate
the scattering vector ${\bf Q}$ and the vector ${\bf q}^*$ on the
FS.  In the case of the full self-energy, Eq.~(\ref{n4}), the
energy-momentum conservation law for three quasipartricles does
not impose such strong limitations.

\subsection{Resonance  mode}

Experimentally, the RM energy $E_{\rm r}$ decreases with
underdoping following the superconducting transition temperature,
$E_{\rm r} \simeq 5.3 k_{\rm B} T_{\rm c}$, but only weakly
depends on temperature (see, e.g., Refs.~\cite{Sidis04,
Eschrig06}). Now we discuss the temperature and doping dependence
of the RM and its dispersion within our theory  for the $d$-wave
pairing.
\par
\begin{figure}
\resizebox{0.4\textwidth}{!}{%
  \includegraphics{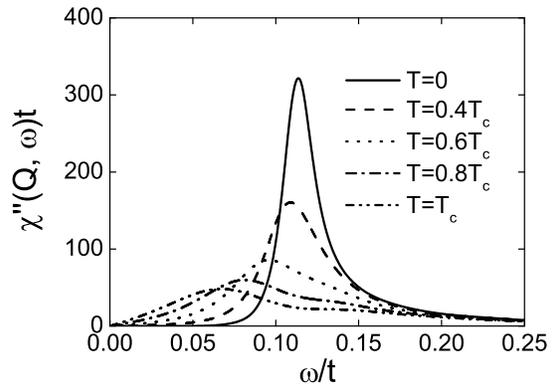}}
\caption{Temperature dependence of the spectral function
$\chi''({\bf Q}, \omega)$ at $\delta=0.2$.}
 \label{Fig_6}
\end{figure}
\begin{figure}
\resizebox{0.4\textwidth}{!}{%
  \includegraphics{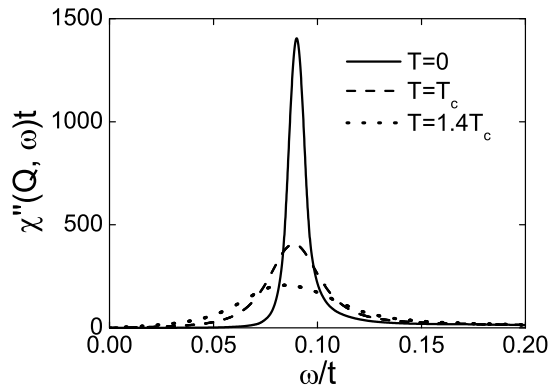}}
\caption{Temperature dependence of the spectral function
$\chi''({\bf Q}, \omega)$ at $\delta=0.09$.}
 \label{Fig_7}
\end{figure}

The temperature dependence of the spectral function in the
overdoped case  $\delta = 0.2$ is shown in Fig.~\ref{Fig_6}. It
has high intensity at low temperatures, but strongly decreases
with temperature and becomes very broad at $T \sim T_{\rm c}$ as
found in experiments (see Ref.~\cite{Bourges98}). In
Fig.~\ref{Fig_7} the temperature dependence of the spectral
function  for the underdoped case  $\delta = 0.09$ is plotted.
Whereas the resonance energy decreases with underdoping, the
intensity of the RM greatly increases in accordance with
experiments. The RM energy weakly depends on temperature and is
still quite visible at $T = T_{\rm c}$ and even at $T = 1.4
T_{\rm c}$.
\par
\begin{figure}
\resizebox{0.39\textwidth}{!}{%
  \includegraphics{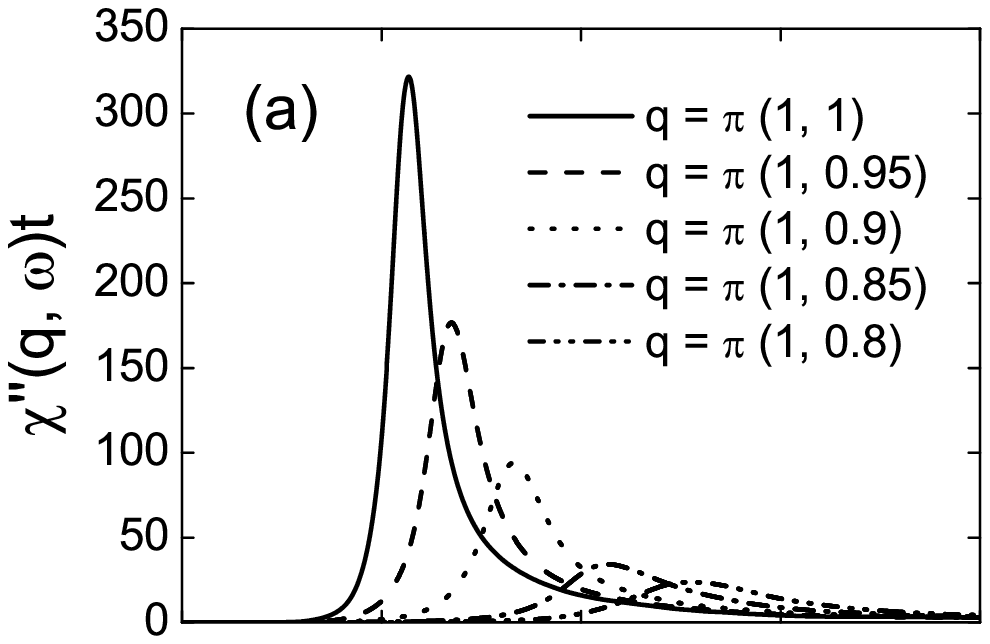}}\\
\mbox{$\;$}  \resizebox{0.4\textwidth}{!}{%
  \includegraphics{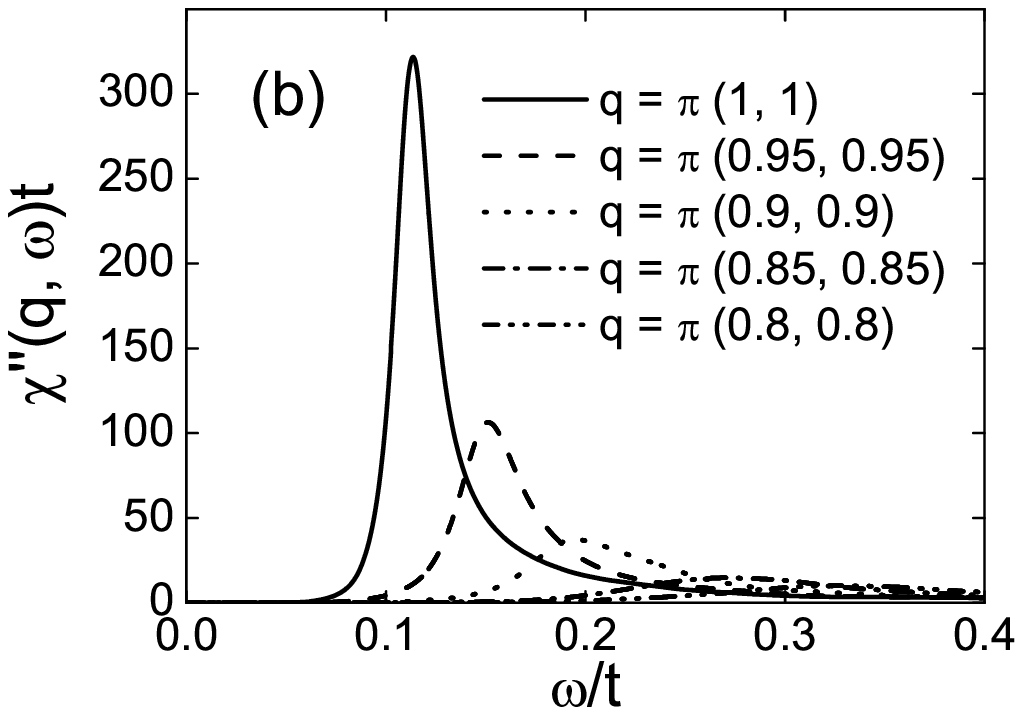}}
\caption{Spectral function $\chi''({\bf q}, \omega)$ for the wave
vectors  (a)  ${\bf q} = \pi (1, \xi)$ and (b)  ${\bf q} = \pi
(\xi, \xi)$ at  $T=0$ for $\delta=0.2$.}
 \label{Fig_8}
\end{figure}
\begin{figure}
\resizebox{0.4\textwidth}{!}{%
\includegraphics{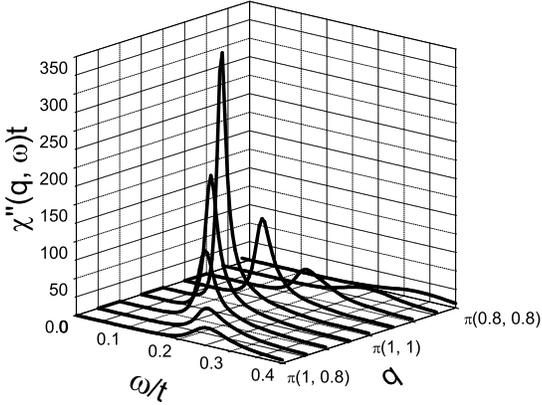}}
\caption{Spectral function $\chi''({\bf q}, \omega)$ near the wave
vector    ${\bf Q} = \pi(1,1)$  at $T=0$ for $\delta=0.2$.}
 \label{Fig_9}
\end{figure}

The dispersion of the spectral function for $\delta = 0.2$ is
shown in Figs.~\ref{Fig_8} and  \ref{Fig_9}. A strong suppression
of the spectral-function intensity away from  ${\bf Q} =
\pi(1,1)$ even at $T=0$  explains the resonance-type behavior of
the function at low temperatures. This suppression of the
intensity is in accord with the  sharp increase of the damping
away from ${\bf Q}$ shown in  Fig.~\ref{Fig_3}.
As can be seen from Figs.~\ref{Fig_8} and ~\ref{Fig_9}, we obtain
an upward dispersion of the resonance energy $\Omega({\bf Q + q})
$. Our numerical results may be well fitted by the quadratic
dispersion law for small wave vectors, $(q_x, q_y) \lesssim 0.2
\pi$ ,
\begin{equation}
\Omega({\bf Q + q}) = [E_{\bf r}^2 + c^2 (q_x^2 + q_y^2)]^{1/2},
 \label{n5}
\end{equation}
where $E_{\bf r}/t = 0.12 \, (0.09)$ and $c/a\,t  = 0.36\,(0.45)$
for $\delta = 0.2 \,(0.09 )$. In the conventional units we obtain
$E_{\bf r} = 38$~meV,  $c = 425$~meV$\cdot$\AA~  for $\delta =
0.2$ and $E_{\bf r} = 28$~meV,  $c = 545$~meV$\cdot$\AA~  for
$\delta = 0.09$, where we take $a = 3.82$~\AA~ and $t = 313$~meV.
Note that the quadratic dispersion  was reported in several
papers. For example, the dispersion (\ref{n5}) was found  in
Ref.~\cite{Stock05} for the acoustic mode in YBCO$_{6.5}$
($\delta = 0.09$) with the parameters $E_{\bf r} = 33$~meV and $c
\simeq 360$~meV$\cdot$\AA~ that qualitatively agrees with our
results.
\par
We have not found the  downward dispersion at energies below the
RM detected in neutron-scattering experiments, for example,  on
YBCO$_{6.5}$ (Ref.~\cite{Stock05}) and on YBCO$_{6.6}$
(Refs.~\cite{Hinkov07,Hinkov10}). However, as  argued in
Ref.~\cite{Stock05}, two distinct regions of  spin excitations
may be suggested: a low-energy part below $E_{\rm r} \approx
33$~meV, which can be described as incommensurate stripe-like
collective spin excitations for the acoustic mode, and a
high-energy part, which has a spin-wave character.  The
high-energy part of the spin excitations has an isotropic
in-plane dispersion while  the low-energy excitations show a
one-dimensional character.  The different nature of the two parts
of the spectrum is also revealed in their temperature dependence:
the low-energy acoustic part of the spectrum is strongly
influenced by the superconducting transition, while the
high-energy part of the spectrum does not change appreciably with
temperature up to $85$~K. Similar differences between the upward
and downward dispersions were found in Refs.~\cite{Hinkov07} and
\cite{Hinkov10}.
\par
The downward dispersion of the lower-energy part of the spectrum
was explained within the stripe-like models, as discussed in the
Introduction. It  was also found for a Fermi liquid model in the
RPA approach (see, e.g., Ref.~\cite{Eremin05}) or, for the
$t$-$J$ model, within the particle-hole bubble approximation (see
Refs.~\cite{Onufrieva02,Sherman03,Sega06}). The dispersion was
explained by a special wave-vector dependence of the
particle-hole bubble diagram  related to the wave-vector
dependence of the $d_{x^2 - y^2 }$ superconducting gap and to a
specific for cuprates 2D FS. Since in our theory beyond the RPA
the RM energy does not critically depend on specific properties
of the FS and the superconducting gap, the downward dispersion
cannot be found. To discuss this problem  in detail, the role of
stripe excitations in the spin-excitation spectrum should be
elucidated.
\par
\begin{figure}
\resizebox{0.4\textwidth}{!}{%
 \includegraphics{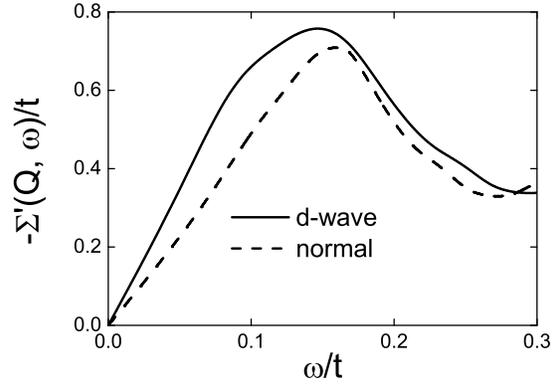}}
\caption{Real part of the self-energy $\, \Sigma'({\bf Q},
\omega)\,$  for the $d$-wave pairing in comparison with the
normal state at $T=0 $ for $\delta=0.2$.}
 \label{Fig_10}
\end{figure}
\begin{figure}
\resizebox{0.4\textwidth}{!}{%
 \includegraphics{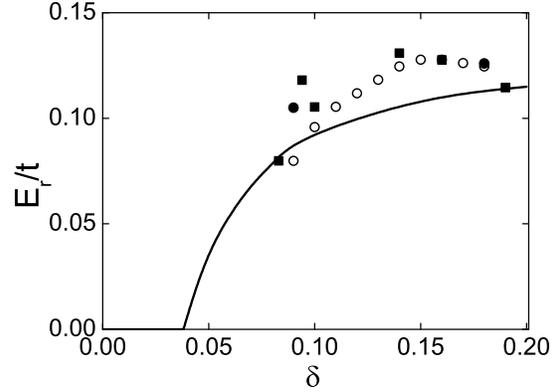}}
\caption{Energy of the resonance mode $E_{\rm r}$ at $T=0$  as a
function of hole doping $\delta$ in comparison with  experimental
data for YBCO  from Ref.~\cite{Sidis04} (open circles),
Ref.~\cite{Eschrig06} (full squares), and Refs.~\cite{Bourges98},
\cite{Stock04} (full circles).}
 \label{Fig_11}
\end{figure}
Now we discuss the doping dependence of  the RM energy $E_{\rm
r}(\delta)$.  At low temperatures, the real part of the
self-energy $\, \Sigma'({\bf Q}, \omega) < 0\,$ is quite large as
shown in Fig.~\ref{Fig_10}. This considerably softens  the energy
of spin excitations  $\omega_{\bf Q}$ in the GMFA,
Eq.~(\ref{b7}), shifting the pole of the spectral function
$\chi''({\bf Q}, \omega)$ to a lower energy:
$\,\widetilde{\omega}_{\bf Q} = [\omega_{\bf Q}^2 -
\widetilde{\omega}_{\bf Q} \, |\Sigma'({\bf Q},
\widetilde{\omega}_{\bf Q})|]^{1/2} \,$. Experimentally, the RM
energy is measured by the position $E_{\rm r}$ of the maximum in
the spectral function $\chi''({\bf Q}, \omega)$ which deviates
from  $\,\widetilde{\omega}_{\bf Q} $ due to a finite width of
excitations. In Fig.~\ref{Fig_11}   the doping dependence $E_{\rm
r}(\delta)$ in the superconducting state at $ T=0 $ determined by
the maximum of the spectral function is plotted. Thereby,  for the
doping dependence of $\Delta_0(\delta) = 1.76 \, T_{\rm
c}(\delta)$  we used  the universal empirical formula  $\, T_{\rm
c}(\delta) = T_{\rm c, max}\, [ 1 - \beta\,(\delta - \delta_{\rm
opt})^2 ] \,$~\cite{Tallon95},  where $\, \delta_{\rm opt}\ =
0.16, \, T_{\rm c, max} = 93$~K, and the value of  $\, \beta =
75\, $ was fitted to obtain $T_{\rm c} = 59$~K for $ \,\delta =
0.09 $ in YBCO$_{6.5}$, Ref.~\cite{Stock04}. With decreasing
$\delta$, $E_{\rm r}$ decreases which qualitatively agrees with
the experimental data.  The energy of the RM tends to zero at the
critical doping $\delta_c = 0.038$ below which the long-range AF
order emerges at $T = 0$, as we have shown in
Ref.~\cite{Vladimirov09}. So, in our scenario the RM is just the
soft mode which brings about the long-range AF order below the
critical doping.
\par
Experimentally, in the overdoped region the RM energy decreases
with increasing doping (see  Fig.~\ref{Fig_11}  and, e.g.,
Ref.~\cite{Pailhes06}), while in our theory  $E_{\rm r}$ tends to
increase due to the  increasing energy $\widetilde{\omega}_{\bf
Q}$. The agreement with experiments in Fig.~\ref{Fig_11}  can be
improved in the optimally doped region,  $\delta \lesssim 0.2 $,
if we change the parameter $t$  to $t = 0.4$~eV
instead of the adopted value  $t = 0.313$~eV. We also note that in the
approach of Ref.~\cite{Sega03} a too large superconducting gap
$\Delta_0 \sim 0.1 t$  (in comparison with our value $\Delta_0 \sim
0.044\, t$) has to be taken to fit the RM energy to the
experimentally observed one.
\par
\begin{figure}
\resizebox{0.4\textwidth}{!}{%
  \includegraphics{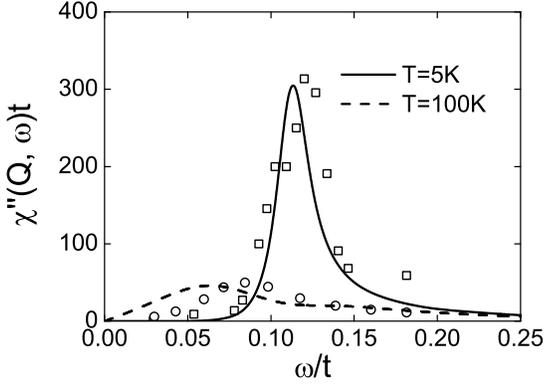}}
\caption{Spectral function $\chi''({\bf Q}, \omega)$ for doping
$\delta = 0.2$ compared to experimental data  for YBCO$_{6.92}$,
Ref.~\cite{Bourges98}, at $T=5K$ (squares) and $T=100K$
(circles).}
 \label{Fig_12}
\end{figure}
\begin{figure}
\resizebox{0.4\textwidth}{!}{%
  \includegraphics{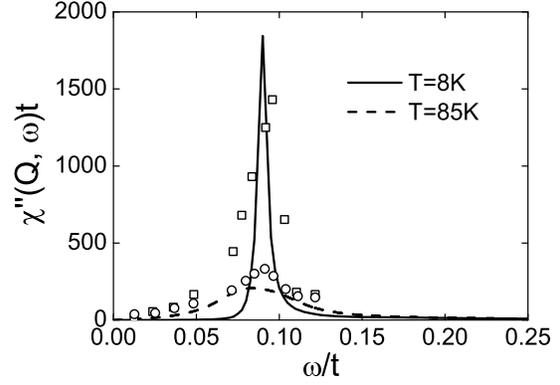}}
\caption{Spectral function $\chi''({\bf Q}, \omega)$ for doping
$\delta = 0.09$ compared to experimental data  for YBCO$_{6.5}$,
Ref.~\cite{Stock04},  at $T=8K$ (squares) and $T=85K$ (circles).}
 \label{Fig_13}
\end{figure}
In Fig.~\ref{Fig_12}  we compare our results with the
neutron-scattering data for the nearly optimally doped
YBCO$_{6.92}$ single crystal~\cite{Bourges98} at $T=5K$  and
$T=100K$. In this sample, $T_{\rm c} = 91$~K and the RM energy
$E_{\rm r} \simeq 40$~meV$ = 5.1 k_{\rm B}\, T_{\rm c} > 2\,
\Delta_0 $ (taking $2\Delta_0(\delta) = 3.52\,k_{\rm B} T_{\rm
c}(\delta)$ we have  $E_{\rm r} \simeq 2.9 \Delta_0 $). For
$\delta = 0.2$,  our calculations yield $E_{\rm r} = 0.12 t =
38$~meV$= 4.8 k_{\rm B}\, T_{\rm c} = 2.7\, \Delta_0 $ ($t =
0.313$~eV, $k_{\rm B}\, T_{\rm c} = 0.025 t$; see Sec.~III.A).
Concerning the heights of the peaks, the scale of the
experimental data, given in arbitrary units, was adjusted by
fitting only the peak height at $T=5K$  to our result at the same
temperature.
\par
In Fig.~\ref{Fig_13}  our results are compared with the
experimental data for  the underdoped ortho-II YBCO$_{6.5}$
single crystal with $E_{\rm r} = 33$~meV$=6.5 \,k_{\rm B}T_{\rm
c}= 3.7\, \Delta_0 $ at $T=8K$ and $T=85K$ (see Fig. 14 in
Ref.~\cite{Stock04}). For $\delta = 0.09$,  our theory gives
$E_{\rm r} = 0.09 t = 28$~meV$ = 5.6 \, k_{\rm B}\, T_{\rm c} =
3.2\, \Delta_0 $. Here,  the experimental peak heights were  also
scaled by fitting only the peak height at $T=8K$  to our result
at the same temperature. We note a weak temperature dependence of
the RM energy observed experimentally and obtained in our
calculation. In both compounds the RM energy is larger than the
superconducting excitation energy, $ 2 \Delta_0$, while in the
spin-1 exciton scenario the RM energy $E_{\rm r}$ has to be less
than $ 2 \Delta_0$.  So we obtain a good agreement of our theory
with neutron-scattering experiments on YBCO crystals both near
the optimal doping and in the underdoped region.
\par
It is also interesting  to compare our results with the recent
detailed  experimental study of the spin-excitation spectrum
reported in Refs.~\cite{Hinkov07} and ~\cite{Hinkov10} for the
twin-free YBCO$_{6.6}$ crystal with $T_{\rm c} = 61$~K at hole
doping $\delta = 0.12$. For that, we calculate the spectral
function $\chi''({\bf q}, \omega)$ for the same hole doping and
$T_{\rm c}$. From the experimental value $\omega_{\rm r}  =
38$~meV and  damping $\Gamma = 11$~meV we calculate $E_{\rm r} =
36.4$~meV~\cite{Er}. To get agreement with the theoretical value
$E_{\rm r} = 0.091 t $ we adopt $t = 0.4$~eV, as also inferred
from the ARPES data in YBCO$_{6.6}$.~\cite{Dahm09} For the upper
branch of spin excitations above the RM we obtain a quadratic
dispersion  given by Eq.~(\ref{n5}), whereas
in~Ref.~\cite{Hinkov10}, Eq.(11), the dispersion was fitted by a
quartic power law.
\par
\begin{figure}[ht!]
\resizebox{0.4\textwidth}{!}{%
 \includegraphics{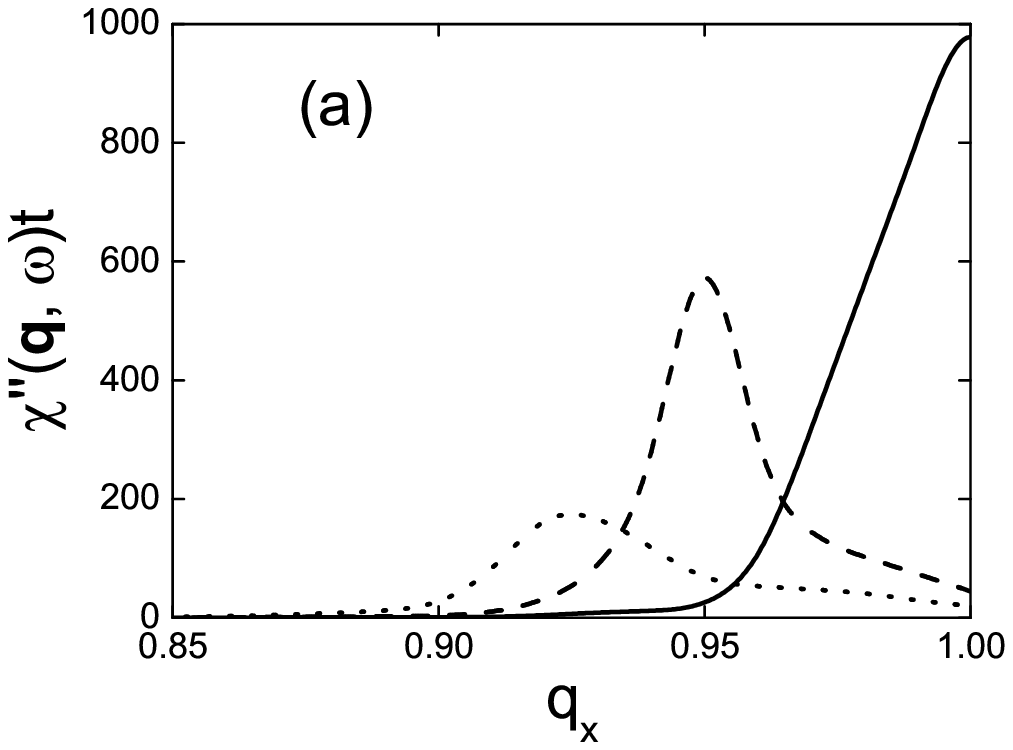}}
\resizebox{0.4\textwidth}{!}{%
 \includegraphics{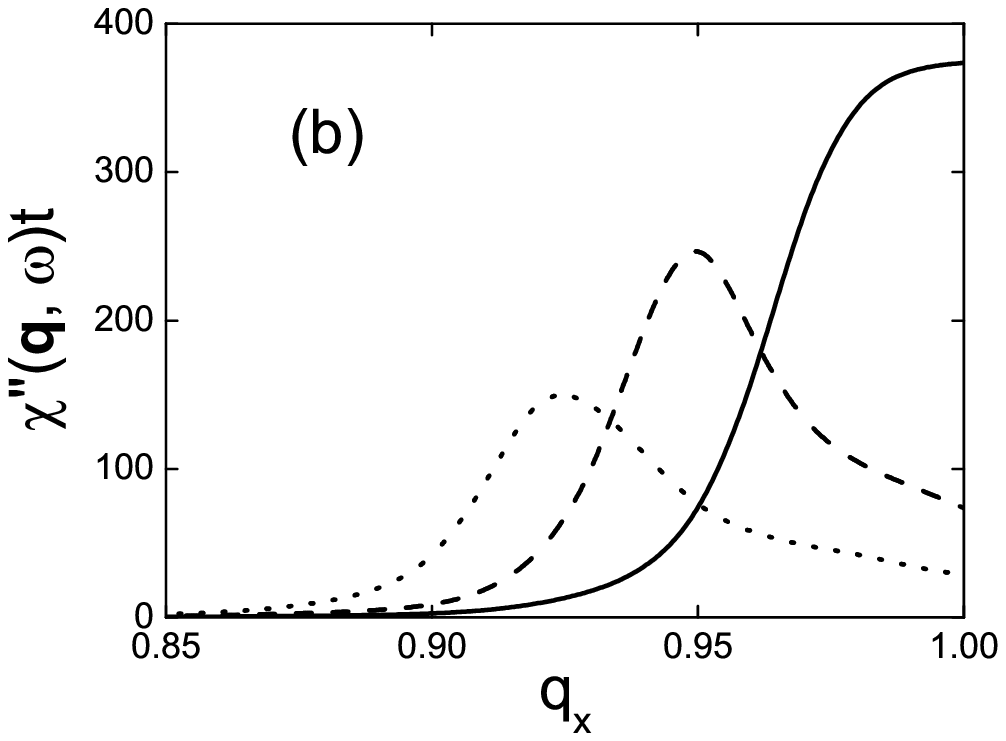}}
\caption{Spectral function   $\chi''({\bf q }, \omega)\,$ for
${\bf q } = \pi (q_x, 1)$ at doping $\delta = 0.12$ and energies
$\omega/E_{\rm r} = 1$~(bold line), 1.24~(dashed line),   and
1.58~(dotted line) at (a) $T =0$ and (b) $T = 70$~K (in units of
$t = 0.4$~eV). }
 \label{Fig_14}
\end{figure}
For a more detailed comparison of the RM dispersion with the
experimental data in Ref.~\cite{Hinkov10} for the upper branch, in
Fig.~\ref{Fig_14} we present  the spectral function $\chi''({\bf
q}, \omega)\,$ at  $T =0$  and  $T = 70$~K for the wave vectors
${\bf q } = \pi(q_x, 1)$ at energies $\omega/E_{\rm r} = 1,\,
1.24$,   and 1.58 which, putting $E_{\rm r}= \omega_{\rm r}$,
corresponds to $\omega =  38 $,  47, and 60~meV in
Ref.~\cite{Hinkov10}. The damping of the RM in our theory
strongly depends on ${\bf q}$ as shown in Fig.~\ref{Fig_3},
whereas in Ref.~\cite{Hinkov10} the damping was taken to be
$q$-independent, $\Gamma = 11$~meV. In our calculations  we get a
much smaller damping:  $\Gamma_{\bf Q} = 0.0034 t \simeq 1.4$~meV
at $T = 0$. These differences result in much sharper spectral
functions  in our theory  in comparison with the experimental
data.~\cite{Hinkov10} In particular, the half-width at
half-maximum at ${\bf q =  Q } = \pi(1, 1)$ equals to $\Delta q =
0.021 \pi$  in Fig.~\ref{Fig_14}~(a)  and $\Delta q = 0.038 \pi$
in Fig.~\ref{Fig_14}~(b)  in comparison with the values $\Delta q
\approx 0.2 \pi$ at  $T =5$~K and $\Delta q \approx 0.24 \pi$ at
$T = 70$~K in Fig.~3 of Ref.~\cite{Hinkov10} being  about 7--8
times larger than  our results. We believe that the difference in
the damping may be explained by impurity scattering due to a
disorder in the chains  in the YBCO$_{6.6}$ sample or  by an
inhomogeneity produced by dynamical stripe fluctuations.  The
latter may be also the reason why we have not found a change of
the dispersion topology above $T_{\rm c}$ observed in
Ref.~\cite{Hinkov10} which was related to the appearance of an
electronic liquid-crystal state.

\section{Conclusion}

A detailed study of the  DSS in the superconducting state  has
revealed the important role of the spin-excitation damping in the
RM phenomenon. We have found that the low-temperature damping
essentially depends on the gap $\widetilde{\omega}_{\bf Q} \simeq
E_{\rm r}$ in the spin-excitation spectrum at the AF wave vector
${\bf Q}$, while the opening of a superconducting gap
$2\Delta(T)$ below $T_{\rm c}$ is less important. Since the
energy of the RM $E_{\rm r} \sim 5\,k_{\rm B}  T_{\rm c}$ does
not show temperature dependence, at $T \lesssim T_{\rm c}$, the
spin gap at $E_{\rm r}$  plays the dominant role in the
suppression of damping, since the superconducting energy $2
\Delta(T \lesssim T_{\rm c}) \ll E_{\rm r}$. This follows from
Eq.~(\ref {n4}) for the self-energy, where in the decay of a spin
excitation, besides a particle-hole pair, the  contribution from
an additional  spin excitation is taken into account. This is in
contrast  to the particle-hole bubble approximation in Eq.~(\ref
{b12a}) which  is used in the RPA (see, e.g.,
Refs.~\cite{Onufrieva02,Eremin05}) and in similar approximations
in memory-function theories.~\cite{Sega03,Sega06,Sherman03} In
those approximations, the spin-excitation damping is much larger
in the normal state and reveals  a spin gap only at $T \ll T_{\rm
c}$, where the RM appears  (see Fig.~\ref{Fig_5}). The gapped
spin-excitation spectrum in the full self-energy (\ref {n4})
greatly suppresses the damping at $T = 0$ which results in a
comparable  damping both in the superconducting state (either of
$d$-wave or $s$-wave symmetry) and in the normal state, as
demonstrated in Figs.~\ref{Fig_1} and \ref{Fig_3}. The damping is
strongly decreasing in the underdoped region (see
Figs.~\ref{Fig_2} and \ref{Fig_3}) bringing about a much stronger
RM seen also above $T_{\rm c}$ (Fig.~\ref{Fig_7}).
\par
The weak temperature dependence of the RM in cuprates can be
contrasted with the magnetic RM in  iron-arsenide
superconductors. In particular, a perfect scaling of the RM
energy   with the superconducting gap below $T_{\rm c }= 25$~K
observed in the BaFe$_{1.85}$Co$_{0.15}$As$_2$
crystal~\cite{Inosov10} supports the spin-1 exciton scenario
expected in AF metals where the SDW instability  determines the
spin-excitation spectrum. Contrary to iron-arsenides, strong
electron correlations in  cuprates  lead to a large AF exchange
interaction between localized copper spins  that results in
temperature independent spin-wave-like excitations. A scenario of
preformed Cooper pairs in cuprates, which may also explain a weak
RM temperature dependence,   seems to contradict experiments (for
discussions  see Ref.~\cite{Plakida10}).
\par
With decreasing hole concentration the energy of the RM decreases
and shows the dependence close to that observed in
neutron-scattering experiments, $E_{\rm r} \sim 5 k_{\rm B}T_{\rm
c} $ (see Fig.~\ref{Fig_11}). Due to the important role of gapped
spin excitations in the damping of the RM, its energy $E_{\rm r}$
does not critically depend on the superconducting gap energy $ 2
\Delta(T)$ and hence, on temperature and peculiarities of the
electronic spectrum in cuprates, contrary to the theories based
on the RPA (see, e.g., Refs.~\cite{Onufrieva02,Eremin05}). In
particular, $E_{\rm r}$ is found to  be larger than $ 2 \Delta_0$
as observed in experiments  (see Figs.~\ref{Fig_12} and
\ref{Fig_13}).
\par
It should be stressed, that the damping of the RM in our
microscopic theory within the $t$--$J$ model  is  determined by
the kinematical interaction induced by the kinetic energy $\,t \,$
of electrons  moving in a singly occupied Hubbard subband. This
interaction is absent in conventional fermion models in which
strong electron correlations are neglected and the spin-electron
scattering is determined by a phenomenological interaction with a
coupling constant as a fit parameter.
\par
Finally, let us note that our approach, using  the MFA for the
electronic spectral functions, Eqs.~(\ref{n1a}) and (\ref{n1b}),
in the computation of the self-energy~(\ref{n4}), has to be
considered as a first step towards a fully self-consistent theory
of the DSS in the $t$--$J$ model. However, we believe that the
consideration of  more accurate fermionic GFs in
Eqs.~(\ref{b11a}) and (\ref{b11b}) beyond the Hubbard-I
approximation (see, e.g., Ref.~\cite{Plakida07}) should not
change our main conclusions, since the electron dispersion, as
shown in Fig.~\ref{Fig_2},  do not play an essential role in our
theory  contrary to the spin-exciton scenario.


\acknowledgments

\indent We thank V. Hinkov for fruitful discussions. Partial
financial support by the Heisenberg--Landau Program of JINR is
acknowledged.  One of the authors (N.P.) is grateful to the
MPIPKS, Dresden, for the hospitality  during his stay at the
Institute, where a  part of the present work has been done.

\appendix

\section{Mode-coupling approximation}

To calculate the self-energy  (\ref{b3}) we use   the MCA for the
time-dependent multi-particle correlation function which appears
in   the spectral representation of the relaxation function
\begin{eqnarray}
\Sigma''({\bf q},\omega)&= & - \frac{1}{2 \omega\, m({\bf q})}
[I({\bf q}, -\omega) - I({\bf q}, \omega)],
\nonumber\\
I({\bf q},\omega) &= & \,\int_{-\infty}^{\infty}\,  dt \,
e^{i\omega t}\langle \, \ddot{S}^{-}_{{\bf q}}|
\ddot{S}^{+}_{-{\bf q}} (t)\rangle ^{\rm proper},
 \label{A1}
\end{eqnarray}
where
\begin{equation}
 -\ddot{S}_{i }^{+} =[[{S}_{i}^{+},\, (H_{t} + H_{J})]\, ,
  \, (H_{t} +H_{J} )] \equiv \sum_{\alpha} F_{i}^{\alpha},
     \label{A2}
\end{equation}
determines  the force correlation functions of the force operators
$\, F_{i}^{\alpha}\,$ denoted by the index $\alpha = tt,\, tJ,\,
Jt, \, JJ$. As discussed in Ref.~\cite{Vladimirov09}, at a
sizeable doping $\delta \sim 0.1$  considered in this paper, only
the term $F_{i}^{tt}$ can be taken into account, since  all other
terms give  negligible contributions. For this term we have
\begin{eqnarray}
 F_{i}^{tt} & = &
\sum_{j,n}t_{ij}\Bigl\{ t_{jn}\left[H^{-}_{ijn}+H^{+}_{nji}\right]
  - t_{in}\left[H^{-}_{jin}+H^{+}_{nij}\right]\Bigr\},
 \nonumber \\
H^{\sigma}_{ijn}&=&X_{i}^{\sigma 0}X_{j}^{+-}X_{n}^{0\sigma} +
X_{i}^{+0}(X_{j}^{00}+X_{j}^{\sigma\sigma})X_{n}^{0-} .
 \label{A3}
\end{eqnarray}
Following the reasoning of Ref.~\cite{Vladimirov09}, in Eq.
(\ref{A3}) only products of operators on different sites are taken
into account. After the Fourier transformation to the ${\bf q
}$-space we obtain the  force-force correlation function
\begin{eqnarray}
&& \langle \, [F_{\bf q}^{tt}]^{\dag} \, | \, F_{\bf q}^{tt}(t)
\rangle =(2t)^4 \, \sum_{{\bf q_1, q_2}} \sum_{{\bf q'_1, q'_2}}
\nonumber\\
&& \langle [{\Lambda}_{\bf q'_1, q'_2, q'_3}\,
 H^{-}_{{\bf q'_1, q'_2, q'_3}} +
  {\Lambda}_{\bf q'_3, q'_2, q'_1}
   \,H^{+}_{{\bf q'_1, q'_2, q'_3}}]^{\dag} \,|\quad
\label{A4} \\
&  & [{\Lambda}_{\bf q_1, q_2, q_3} \, H^{-}_{{\bf q_1, q_2, q_3
}}(t) + {\Lambda}_{\bf q_3, q_2, q_1}
  \, H^{+}_{{\bf q_1, q_2, q_3}}(t)] \rangle ,
 \nonumber
\end{eqnarray}
where  ${\bf q}_3 = {\bf q}-{\bf q}_1-{\bf q}_2$ and ${\bf q'}_3
= {\bf q}-{\bf q'}_1-{\bf q'}_2$. Here we introduce the vertex
function
\begin{equation}
 \Lambda_{{\bf q}_1 {\bf q}_2 {\bf q}_3}
= 4(\gamma_{{\bf q}_3+{\bf q}_2}
  -  \gamma_{{\bf q}_1 })\, \gamma_{q_3}
+ \gamma_{{\bf q}_2}-\gamma_{{\bf q}_1 + {\bf q}_3} ,
 \label{A5}
\end{equation}
where the terms linear in $\gamma_{{\bf q}}$ reflect the
exclusion of terms in $ F_{i}^{tt}$ with coinciding sites.
\par
In the MCA we assume that the propagation of electronic-  and
bosonic-type excitations at different lattice sites in
Eq.~(\ref{A3}) occurs independently which results in the
decoupling of the correlation function (\ref{A4}) into the
corresponding single-particle time-dependent correlation
functions. As it turned out by numerical evaluations  (see also
Ref.~\cite{Vladimirov09}),  the contribution from the charge
excitations given  by $(X_{j}^{00}+X_{j}^{\sigma\sigma})$ in
Eq.~(\ref{A3}) can be neglected  in comparison with the
spin-excitation contribution given by $ X_{j}^{+-} = S_{j}^{+} $.
In this approximation we obtain  the spin-diagonal correlation
functions for the normal state
\begin{eqnarray}
&&\langle [H^{\sigma}_{{\bf q'_1, q'_2, q'_3}}]^{\dag}  \,|\,
H^{\sigma}_{{\bf q_1, q_2, q_3 }}(t) \rangle
\nonumber \\
 &=&\langle   X_{\bf q'_3}^{\sigma 0}  S_{\bf -q'_2}^{-}
  X_{\bf q'_1}^{0\sigma} \,| \,
X_{\bf q_1}^{\sigma 0}(t) S_{\bf q_2}^{+}(t)
 X_{\bf q_3}^{0\sigma}(t)\rangle
\nonumber \\
&=& \langle  X_{\bf q_1}^{0\sigma} X_{\bf q_1}^{\sigma 0}(t)
\rangle \,  \langle S_{\bf -q_2}^{-} S_{\bf q_2}^{+}(t)\rangle \,
\langle X_{\bf q_3}^{\sigma 0}  X_{\bf q_3}^{0\sigma}(t)\rangle
 \nonumber \\
&\times& \, \delta_{{\bf q}_1, {\bf q}'_1 }\,\delta_{{\bf q}_2,
{\bf q}'_2}
  \delta_{{\bf q}_3, {\bf q}'_3 }
\label{A6}
\end{eqnarray}
In the superconducting state we additionally  take into account
pair correlation functions which appear in the spin off-diagonal
terms
\begin{eqnarray}
&&\langle [H^{\bar\sigma}_{{\bf q'_1, q'_2, q'_3}}]^{\dag} \,| \,
H^{\sigma}_{{\bf q_1, q_2, q_3 }}(t) \rangle
\nonumber \\
 &=&\langle  X_{\bf q'_3}^{\bar\sigma 0}  S_{\bf -q'_2}^{-}
  X_{\bf q'_1}^{0\bar\sigma}\, |
\, X_{\bf q_1}^{\sigma 0}(t) S_{\bf q_2}^{+}(t)
 X_{\bf q_3}^{0\sigma}(t)\rangle
\nonumber \\
& = & -  \langle  X_{\bf -q_3}^{0{\bar\sigma}} X_{\bf
q_3}^{0\sigma}(t) \rangle \,
  \langle S_{\bf -q_2}^{-} S_{\bf q_2}^{+}(t)\rangle \,
 \langle X_{\bf -q_1}^{{\bar\sigma} 0}  X_{\bf q_1}^{\sigma0}(t)  \rangle
 \nonumber \\
&\times& \, \delta_{{\bf q}_1, -{\bf q}'_3 }\,\delta_{{\bf q}_2,
{\bf q}'_2}  \delta_{{\bf q}_3, -{\bf q}'_1 }.
 \label{A7}
\end{eqnarray}
By substituting  the MCA  correlation functions (\ref{A6}) and
(\ref{A7}) in Eq.~(\ref{A1}) we obtain    the multi-particle
correlation function
\begin{eqnarray}
&&I({\bf q},\omega) =
 \int_{-\infty}^{\infty} dt
e^{i\omega t}\frac{1}{N^2}\sum_{{\bf q_1, q_2}}\,\langle S_{\bf
-q_2}^{-} S_{\bf q_2}^{+}(t)\rangle
\label{A8}\\
&&\times\big[({\Lambda}^2_{\bf q_1, q_2, q_3} + {\Lambda}^2_{\bf
q_3, q_2, q_1}) \langle X_{\bf q_1}^{0\sigma} X_{\bf q_1}^{\sigma
0}(t) \rangle\langle X_{\bf q_3}^{\sigma 0} X_{\bf
q_3}^{0\sigma}(t)\rangle
\nonumber \\
&& -  {\Lambda}_{\bf q_1, q_2, q_3} {\Lambda}_{\bf q_3, q_2, q_1}
 \sum_{\sigma} \langle X_{\bf -q_1}^{\bar{\sigma} 0}
X_{\bf q_1}^{\sigma 0}(t) \rangle
 \langle  X_{\bf -q_3}^{0 \bar{\sigma}} X_{\bf q_3}^{0\sigma}(t)
\rangle \big]. \nonumber
\end{eqnarray}
Using  the spectral representation for the time-dependent
correlation functions (see, e.g.,  Ref.~\cite{Zubarev60})
\begin{equation}
\langle B A(t)\rangle =\int_{-\infty}^{\infty} d\omega
\,e^{-i\omega t}  f(\omega)[-(1/\pi)]{\rm Im} \langle \!\langle
A|B \rangle\!\rangle_{\omega},
 \end{equation}
where $f(\omega)$ is the Fermi (Bose) function $n(\omega)$ (
$N(\omega)$), after integration over time $t$ we derive
Eq.~(\ref{b10}).


\begin{thebibliography}{99}

\bibitem{Bourges98} Ph. Bourges, in  \textit{ The Gap Symmetry and
Fluctuations in High Temperature  Superconductors}, ed. by
J.~Bok, G.~Deutscher, D.~Pavuna and  S.A.~Wolf (Plenum Press,
1998), p. 349-371 (Vol. \textbf{371} in NATO ASI  series,
Physics). [arXiv:cond-mat/9901333].

\bibitem{Sidis04} Y. Sidis, S. Pailh\`{e}s,  B. Keimer,  Ph. Bourges,  C. Ulrich,
and L.P.  Regnault,  phys. stat. sol. (b) \textbf{241}, 1204
(2004).


\bibitem{Sidis07} Y. Sidis, S. Pailh\`{e}s, V. Hinkov, B. Fauqu\'{e},
C. Ulrich, L. Capogna, A. Ivanov, L.-P. Regnault, B. Keimer, and
P. Bourges, C. R. Physique, \textbf{8}, 745 (2007).

\bibitem{Eschrig06}  M.  Eschrig, Adv. Phys. \textbf{55}, 47 (2006).


\bibitem{Plakida10} N.M. Plakida, \textit{High Temperature
Cuprate Superconductors}, (Springer, Berlin Heidelberg New York,
2010), pp.~51--99.


\bibitem{Si93} Q. Si, Y. Zha, K. Levin, and J.P. Lu,
 Phys. Rev. B \textbf{47}, 9055 (1993).



\bibitem{Fong00} H.F. Fong,  Ph. Bourges, Y. Sidis,, L.P. Regnault,
J. Bossy, A. Ivanov, D.L. Milius, I.A. Aksay,  and B. Keimer,
Phys.~Rev.~B  \textbf{61}, 14773 (2000).


\bibitem{Dai01} P. Dai,  H.A. Mook, R.D. Hunt,
F. Do\u{g}an,  Phys.~Rev.~B \textbf{63}, 054525 (2001).


\bibitem{Yamase06}  H. Yamase, and W. Metzner, Phys. Rev. B
\textbf{73}, 214517 (2006).



\bibitem{Rossat91}  J. Rossat-Mignod, L. P. Regnault, C. Vettier,
P. Bourges, P. Burlet, J. Bossy, J. Y. Henry, and G. Lapertot,
Physica C, \textbf{185--189}, 86 (1991).

\bibitem{He02} H. He, P. Bourges, Y. Sidis, C. Ulrich, L. P. Regnault,
S.~Pailh\`{e}s,  N. S. Berzigiarova, N. N. Kolesnikov, and B.
Keimer, Science \textbf{295}, 1045  (2002).



\bibitem{Yu10} G. Yu, Y. Li, E. M. Motoyama, X. Zhao, N. Barisic,
Y. Cho, P. Bourges, K. Hradil, R. A. Mole, and M. Greven,
   Phys. Rev. B \textbf{81}, 064518 (2010).


\bibitem{Wilson06} S. D. Wilson, P. Dai1, S. Li, S. Chi,
H. J. Kang, and J. W. Lynn, Nature \textbf{442}, 59 (2006).


\bibitem{Stock04} C. Stock, W.J.L. Buyers, R. Liang, D. Peets,  Z. Tun, D. Bonn,
W.N. Hardy, and  R.J. Birgeneau, Phys.~Rev.~B \textbf{69}, 014502
(2004).


\bibitem{Stock05} C. Stock, W.J.L. Buyers, R.A. Cowley, P.S. Clegg,
R. Coldea, C.D. Frost,  R. Liang, D. Peets,  D. Bonn, W.N. Hardy,
and  R.J. Birgeneau, Phys.~Rev.~B  \textbf{71}, 024522 (2005).


\bibitem{Hinkov07}  V.~Hinkov, Ph.~Bourges, S.  Pailh\`{e}es,
Y.~Sidis, A.~Ivanov, C.D.~Frost, T.G.~Perring, C.T.~Lin,
D.P.~Chen, B.~Keimer, Nature Phys. \textbf{3}, 780 (2007).


\bibitem{Hinkov10}  V.~Hinkov, B.~Keimer, A.~Ivanov,
Ph.~Bourges, Y.~Sidis, C.D.~Frost,
http://xxx.lanl.gov/arXiv:cond-mat/1006.3278 (2010).


\bibitem{Dahm09}
T. Dahm, V. Hinkov, S. V. Borisenko, A. A. Kordyuk, V. B.
Zabolotnyy, J. Fink, B. B\"{u}chner, D. J. Scalapino, W. Hanke and
B. Keimer, Nature Phys.  \textbf{5}, 217 (2009).


\bibitem{Christensen04}  N.B. Christensen, D.F. McMorrow,
H.M. R$\o$nnow, B. Lake, S.M. Hayden, G. Aeppli, T.G. Perring, M.
Mangkorntong, M. Nohara, and H. Takagi, Phys. Rev. Lett.
\textbf{93},  147002 (2004).


\bibitem{Vignolle07} B. Vignolle, S.M. Hayden, D.F. McMorrow,
H.M. R$\o$nnow, B. Lake, C.D. Frost, and T.G. Perring, Nature
Phys.  {\bf 3}, 163 (2007).


\bibitem{Lipscombe09} O.J. Lipscombe, B. Vignolle, T.G. Perring, C.D. Frost, and S.M.
Hayden, Phys. Rev. Lett. \textbf{102}, 167002 (2009).


\bibitem{Norman00}    M. R. Norman,
 Phys. Rev. B \textbf{61}, 14751 (2000).


\bibitem{Manske01} D. Manske, I. Eremin, and K. H. Bennemann,
Phys. Rev. B \textbf{63}, 054517 (2001).


\bibitem{Eremin05} I. Eremin, D.K. Morr, A.V. Chubukov,
K. Bennemann, and M.R. Norman,   Phys. Rev. Lett. \textbf{94},
147001 (2005).


\bibitem{Si92} Q. Si,  J.P. Lu, and K. Levin,
 Phys. Rev. B \textbf{45}, 4930 (1992).


\bibitem{Zha93} Y. Zha,  K. Levin,  and Q. Si,
 Phys. Rev. B \textbf{47}, 9124 (1993).



\bibitem{Liu95}  D. Z. Liu, Y. Zha, and K. Levin,
Phys. Rev. Lett. \textbf{75}, 4130 (1995).



\bibitem{Kao00} Y.J. Kao,  Q. Si, and K. Levin,
Phys. Rev. B \textbf{61}, R11898 (2000).


\bibitem{Eremin07} I. Eremin, D.K. Morr, A.V. Chubukov, K. Bennemann,
 Phys. Rev. B \textbf{75},  184534 (2007).


\bibitem{Fauque07} B. Fauqu\'{e}, Y. Sidis, L. Capogna, A. Ivanov,
K. Hradil, C. Ulrich, A.I. Rykov, B. Keimer, and P. Bourges,
Phys. Rev. B \textbf{76}, 214512 (2007).


\bibitem{Pailhes06} S. Pailh\'{e}s, C. Ulrich, B. Fauqu\'{e}, V. Hinkov, Y. Sidis,
A. Ivanov, C.T. Lin, B. Keimer, and P. Bourges, Phys. Rev. Lett.
\textbf{96}, 257001 (2006).


\bibitem{Ogata08} M. Ogata and
H. Fukuyama,  Rep. Prog. Phys. \textbf{71}, 036501  (2008).


\bibitem{Brinckmann02}  J. Brinckmann and P.A. Lee,
Phys. Rev. B \textbf{65}, 014502 (2001).


\bibitem{Hubbard65}  J. Hubbard,
Proc. Roy. Soc. A \textbf{285}, 542 (1965).


\bibitem{Izyumov91} Yu.A. Izyumov and  B.M. Letfulov,
J. Phys.: Condens. Matter \textbf{3} 5373 (1991).



\bibitem{Onufrieva02} F.~Onufrieva and P.~Pfeuty,
 Phys. Rev. B \textbf{65}, 054515 (2002).


\bibitem{Mori65} H. Mori, Prog. Theor. Phys. \textbf{34},  399 (1965).


\bibitem{Sega03}  I. Sega,  P. Prelov\v{s}ek, and J. Bon\v{c}a,
  Phys. Rev. B  \textbf{68}, 054524, (2003).


\bibitem{Prelovsek04} P. Prelov\v{s}ek, I. Sega, and J. Bon\v{c}a,
Phys. Rev. Lett. \textbf{92}, 027002 (2004).

\bibitem{Sega06}    I. Sega  and  P. Prelov\v{s}ek
Phys. Rev. B \textbf{73}, 092516 (2006).

\bibitem{Prelovsek06}  P. Prelov\v{s}ek and I. Sega,
Phys. Rev. B  \textbf{74}, 214501 (2006).

\bibitem{Sherman03} A. Sherman and M. Schreiber,
Phys. Rev. B  \textbf{68}, 094519 (2003).

\bibitem{Sherman06} A.~Sherman  and M.~Schreiber, Fiz. Nizk. Temp.
(Low Temp. Phys., Ukraine) \textbf{32}, 499 (2006).

\bibitem{Vladimirov05} A.A. Vladimirov, D. Ihle, and N. M. Plakida,
Theor. Math. Phys. \textbf{145}, 1576 (2005).

\bibitem{Vladimirov09} A.A. Vladimirov, D. Ihle, and N. M. Plakida,
Phys. Rev. B \textbf{80}, 104425 (2009).

\bibitem{Zeyher10}  R. Zeyher,
Europ. Phys. Lett., \textbf{90}, 17006 (2010).

\bibitem{Tranquada07} J.M. Tranquada, in
{\it Handbook of High-Temperature  Superconductivity. Theory and
Experiment}, ed. by J.R.~Schrieffer  and J.S.~Brooks (Springer,
New York, 2007), pp. 257--298.


\bibitem{Kivelson03}  S.A. Kivelson, E. Fradkin,
V. Oganesyan,  I.P. Bindloss, J.M. Tranquada, A. Kapitulnik, and
 C.  Howald, Rev. Mod. Phys. \textbf{75}, 1201 (2003).


\bibitem{Vojta09} M. Vojta, Adv. Phys. \textbf{58}, 699 (2009).


\bibitem{Tranquada04} J. M. Tranquada, H. Woo, T. G. Perring,
H. Goka, G. D. Gu1, G. Xu, M. Fujita, and  K. Yamada, Nature
 \textbf{429}, 534 (2004).


\bibitem{Vojta04}  M. Vojta and T. Ulbricht,
Phys. Rev. Lett. \textbf{93}, 127002 (2004).


\bibitem{Uhrig04}  G. S. Uhrig, K. P. Schmidt, and
M. Gr\"{u}ninger,  Phys. Rev. Lett. \textbf{93}, 267003 (2004).


\bibitem{Seibold05} G. Seibold and J. Lorenzana,
  Phys. Rev. Lett. \textbf{94},  107006  (2005).


\bibitem{Andersen05} B. M. Andersen and P. Hedeg\"{a}rd,
 Phys. Rev. Lett. \textbf{95},  037002 (2005).


\bibitem{Andersen07} B. M. Andersen and O.F. Sylju{\aa}åsen
 Phys. Rev. B \textbf{75}, 012506  (2007).


\bibitem{Konik08}  R. M. Konik, F. H. L. Essler, and
A. M. Tsvelik,    Phys. Rev. B \textbf{78}, 214509 (2008).


\bibitem{Hinkov04}
V. Hinkov, S. Pailh\`{e}es, Ph. Bourges, Y. Sidis, A. Ivanov,
A.~Kulakov, C.T. Lin,  D.P. Chen,  C. Bernhard, and B. Keimer,
Nature  \textbf{430}, 650 (2004).


\bibitem{Mook00}  H.A. Mook,
P. Dai,  F. Do\u{g}an, and  R.D. Hunt, Nature  \textbf{404}, 729
(2000).


\bibitem{Zubarev60}  D.N. Zubarev, Sov. Phys. Uspekhi \textbf{3}, 320 (1960).

\bibitem{Tserkovnikov81} Yu.A. Tserkovnikov,  Theor. Math. Phys.
\textbf{49}, 993 (1981);  Theor. Math. Phys. \textbf{52}, 712
(1982).

\bibitem{Won94} H.~Won and K. Maki,  Phys. Rev. B
\textbf{49}, 1397 (1994).

\bibitem{Tallon95} J.L. Tallon, C. Bernhard,  H. Shaked,
R.L. Hitterman, and J.D. Jorgensen,    Phys.~Rev.~B \textbf{51},
12911 (1995).

\bibitem{Er} The maximum of the DSS, Eq.~(\ref{b4}),
which defines the RM energy $E_{\rm r}$, is given by the equation
$\, E_{\rm r} = (\omega_{\rm r} /\sqrt{3})[ 1- 2\gamma^2
 + 2\sqrt{1- \gamma^2}]^{1/2},$ where $\gamma =
 \Gamma/\omega_{\rm r}$ and, in our
notation, $\omega_{\rm r}= \widetilde{\omega}_{\bf Q}$ and $2
\Gamma = - \Sigma{''}({\bf Q},\widetilde{\omega}_{\bf Q})$.

\bibitem{Inosov10} D. S. Inosov, J. T. Park, P. Bourges, D. L. Sun, Y. Sidis, A.
Schneidewind, K. Hradil, D. Haug, C. T. Lin, B. Keimer, and V.
Hinkov,  Nature Phys. \textbf{6}, 178 (2010).

\bibitem{Plakida07}   N.M.~Plakida  and  V.S.~Oudovenko,
Journ. of  Exp. and Theor. Phys. \textbf{104}, 230 (2007).



\end{thebibliography}
\end{document}